\documentclass[aps, prl, twocolumn, times, superscriptaddress]{revtex4-1}
\usepackage{bm, amsmath, amsfonts, amssymb}
\usepackage{times}
\usepackage{braket}
\usepackage[dvipdfmx]{graphicx}
\usepackage{float}
\usepackage{color}
\usepackage{comment}

\usepackage{xspace}
\usepackage{stmaryrd}
\usepackage{ulem}

\newcommand{\ii}{\text{i}}

\begin{document}

\title{Topological unification of time-reversal and particle-hole symmetries in non-Hermitian physics}

\author{Kohei Kawabata}
\email{kawabata@cat.phys.s.u-tokyo.ac.jp}
\affiliation{Department of Physics, University of Tokyo, 7-3-1 Hongo, Bunkyo-ku, Tokyo
113-0033, Japan}

\author{Sho Higashikawa}
\affiliation{Department of Physics, University of Tokyo, 7-3-1 Hongo, Bunkyo-ku, Tokyo
113-0033, Japan}

\author{Zongping Gong}
\affiliation{Department of Physics, University of Tokyo, 7-3-1 Hongo, Bunkyo-ku, Tokyo
113-0033, Japan}

\author{Yuto Ashida}
\affiliation{Department of Physics, University of Tokyo, 7-3-1 Hongo, Bunkyo-ku, Tokyo
113-0033, Japan}

\author{Masahito Ueda}
\affiliation{Department of Physics, University of Tokyo, 7-3-1 Hongo, Bunkyo-ku, Tokyo
113-0033, Japan}
\affiliation{RIKEN Center for Emergent Matter Science (CEMS), Wako, Saitama 351-0198, Japan}

\date{\today}

\begin{abstract}
Topological phases are enriched in non-equilibrium open systems effectively described by non-Hermitian Hamiltonians. While several properties unique to non-Hermitian topological systems were uncovered, the fundamental role of symmetry in non-Hermitian physics has yet to be fully understood, and it has remained unclear how symmetry protects non-Hermitian topological phases. Here we show that two fundamental anti-unitary symmetries, time-reversal and particle-hole symmetries, are topologically equivalent in the complex energy plane and hence unified in non-Hermitian physics. A striking consequence of this symmetry unification is the emergence of unique non-equilibrium topological phases that have no counterparts in Hermitian systems. We illustrate this by presenting a non-Hermitian counterpart of the Majorana chain in an insulator with time-reversal symmetry and that of the quantum spin Hall insulator in a superconductor with particle-hole symmetry. Our work establishes a fundamental symmetry principle in non-Hermitian physics and paves the way towards a unified framework for non-equilibrium topological phases.
\end{abstract}

\maketitle

It is Wigner who showed that all symmetries are either unitary or anti-unitary and identified the fundamental role of time-reversal symmetry in anti-unitary operations~\cite{Wigner}. Time-reversal symmetry is complemented by particle-hole and chiral symmetries, culminating in the Altland-Zirnbauer (AZ) ten-fold classification~\cite{AZ-97}. The AZ classification plays a key role in characterizing topological phases~\cite{Kane-review, Zhang-review, Schnyder-Ryu-review} of condensed matter such as insulators~\cite{TKNN-82, Haldane-88, Kane-Mele-05, BHZ-06, Moore-07, Fu-07, Konig-07} and superconductors~\cite{Read-00, Kitaev-01, Ivanov-01, Fu-08}, as well as photonic systems~\cite{Lu-review} and ultracold atoms~\cite{Goldman-review}, all of which are classified into the periodic table~\cite{Schnyder-08, Kitaev-09, Ryu-10, Teo-10}. Whereas the topological phase in the quantum Hall insulator is free from any symmetry constraint and breaks down in the presence of time-reversal symmetry~\cite{TKNN-82, Haldane-88}, certain topological phases are protected by symmetry: for example, the quantum spin Hall insulator is protected by time-reversal symmetry~\cite{Kane-Mele-05, Moore-07, Fu-07, BHZ-06} and the Majorana chain is protected by particle-hole symmetry~\cite{Kitaev-01}.

Despite its enormous success, the existing framework for topological phases mainly concerns equilibrium closed systems. Meanwhile, there has been growing interest in non-equilibrium open topological systems, especially non-Hermitian topological systems~\cite{Rudner-09, Zeuner-15, Weimann-17, Hu-11, Esaki-11, Malzard-15, SanJose-16, Lee-16, Xu-17, Leykam-17, Obuse-17, St-Jean-17, Shen-18, Zhou-18, Parto-18, Segev-18, KK-tsc18, Kunst-18, Yao-18, Kozii-17}. In general, non-Hermiticity arises from the presence of energy or particle exchanges with an environment~\cite{Konotop-review, Christodoulides-review}, and a number of phenomena and functionalities unique to non-conservative systems have been theoretically predicted~\cite{Hatano-96, Bender-98, Bender-07, Makris-08, Klaiman-08, Lin-11, Brody-12, Ge-17, Ashida-17, KK-17, Lau-18, Nakagawa-18} and experimentally observed~\cite{Guo-09, Ruter-10, Regensburger-12, Peng-14, Feng-14, Hodaei-14, Zhen-15, Gao-15, Peng-16, Hodaei-17, Chen-17}. Here symmetry again plays a key role; for example, spectra of non-Hermitian Hamiltonians can be entirely real in the presence of parity-time symmetry~\cite{Bender-98}. Recently, a topological band theory for non-Hermitian Hamiltonians was developed and the topological phase in the quantum Hall insulator was shown to persist even in the presence of non-Hermiticity~\cite{Shen-18}. Moreover, topological lasers were proposed and realized on the basis of the interplay between non-Hermiticity and topology~\cite{St-Jean-17, Parto-18, Segev-18}. However, it has yet to be understood how symmetry constrains non-Hermitian systems in general and how symmetry protects non-Hermitian topological phases.

Here we point out that two fundamental anti-unitary symmetries, time-reversal symmetry and particle-hole symmetry, are two sides of the same symmetry in non-Hermitian physics. In fact, once we lift the Hermiticity constraint on the Hamiltonian $H$, the Wigner theorem dictates that an anti-unitary operator ${\cal A}$ is only required to satisfy
\begin{equation}
{\cal A} H {\cal A}^{-1} = e^{\ii \varphi} H\quad\left( 0 \leq \varphi < 2\pi \right).
	\label{symmetry}
\end{equation}
This suggests that time-reversal symmetry ($\varphi = 0$) and particle-hole symmetry ($\varphi = \pi$) can be continuously transformed to each other in the complex energy plane. This topological unification leads to striking predictions about topological phenomena. In particular, properties intrinsic to topological insulators can appear also in the corresponding topological superconductors, and vice versa: a counterpart of the Majorana chain in a non-Hermitian insulator with time-reversal symmetry and that of the quantum spin Hall insulator in a non-Hermitian superconductor with particle-hole symmetry. We emphasize that such topological phases are absent in Hermitian systems; non-Hermiticity alters the topological classification in a fundamental manner, and non-equilibrium topological phases unique to non-Hermitian systems emerge as a result of the topological unification of time-reversal and particle-hole symmetries.

\begin{figure}[t]
\centering
\includegraphics[width=84mm]{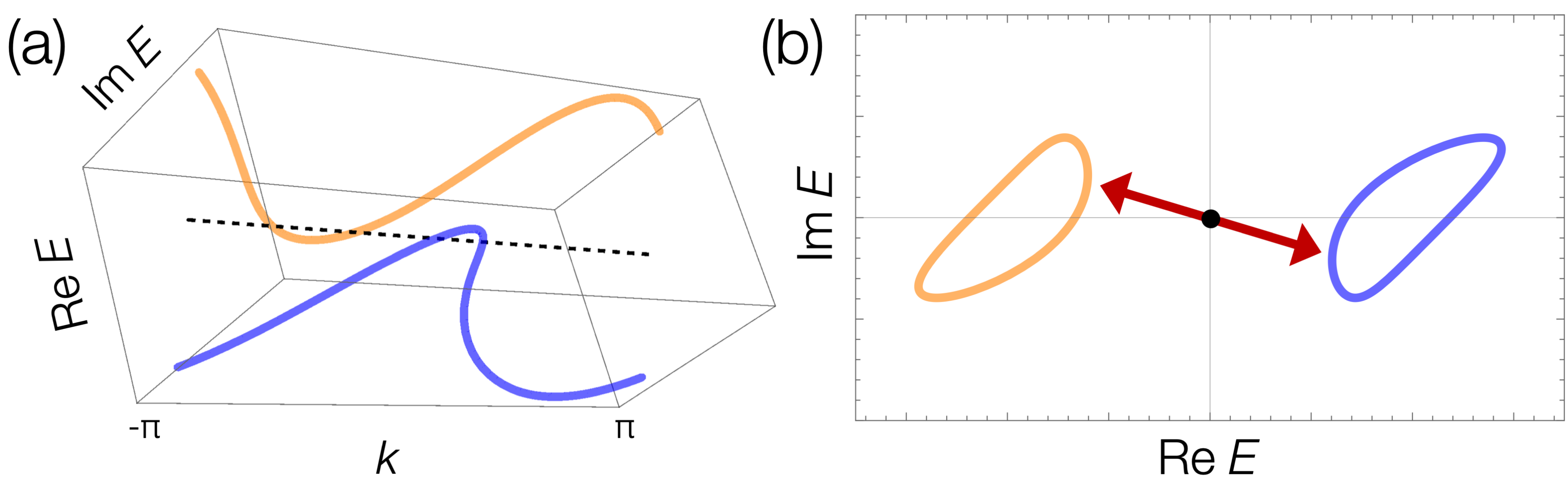} 
\caption{Gapped band structure for a non-Hermitian Hamiltonian. Energy dispersion for two bands (blue and orange curves) in one dimension: (a) $\left( k,\,{\rm Re}\,E,\,{\rm Im}\,E \right)$ with wavenumber $k$ and complex energy $E$, and (b) its projection on the complex energy plane $\left( {\rm Re}\,E,\,{\rm Im}\,E \right)$. The two bands neither touch nor intersect for any $k$, and therefore they are gapped. This definition does not distinguish between real and imaginary parts of energy.}
	\label{fig: gapped complex band}
\end{figure}

\paragraph{Symmetries and complex spectra.\,---}
To go beyond the Hermitian paradigm, it is necessary to revisit some fundamental concepts relevant to topology. We start by defining a gapped complex band. Let us consider a complex-band structure $\{ E_{n} ({\bm k}) \in \mathbb{C} \}$, where ${\bm k}$ is a crystal wavevector in the Brillouin zone and $n$ is a band index. Since a band gap should refer to an energy range in which no states exist, it is reasonable to define a band $n$ to be gapped such that $E_{m} ({\bm k}) \neq E_{n} ({\bm k})$ for all the band indices $m \neq n$ and wavevectors ${\bm k}$ (Fig.~\ref{fig: gapped complex band})~\cite{Shen-18}, which is a natural generalization of the gapped band structure in the Hermitian band theory and explains the experimentally observed topological edge states in non-Hermitian systems~\cite{Zeuner-15, Weimann-17, Obuse-17, St-Jean-17, Parto-18, Segev-18}. Notably, the presence of a complex gap has a significant influence on the non-equilibrium wave dynamics (see Supplementary Note 1 for details). This definition of a complex gap is distinct from that adopted in Ref.~\cite{Gong-18} and hence the corresponding topological classification is different.

\begin{table}[t]
	\centering
	\caption{Constraints on the complex spectra imposed by the Altland-Zirnbauer (AZ) symmetry. In Hermitian systems, time-reversal symmetry places no constraints on the real spectra, while particle-hole symmetry gives zero energy $E=0$ or opposite-sign pairs $\left( E,\,-E \right)$. In non-Hermitian systems, by contrast, time-reversal symmetry gives real energies $E \in \mathbb{R}$ or complex-conjugate pairs $\left( E,\,E^{*} \right)$, while particle-hole symmetry gives pure imaginary energies $E \in \ii \mathbb{R}$ or pairs $\left( E,\,-E^{*} \right)$. Chiral symmetry gives $E=0$ or pairs $\left( E,\,-E \right)$ in both Hermitian and non-Hermitian systems.\\}
	\begin{tabular}{c||c|c} \hline
   		 ~AZ symmetry~ & Hermitian & Non-Hermitian \\ \hline
    		~time-reversal~ & no constraints & ~$E \in \mathbb{R}$ or $\left( E,\,E^{*} \right)$~ \\
		~particle-hole~ & ~~$E=0$ or $\left( E,\,-E \right)$~~ & ~~$E \in \ii \mathbb{R}$ or $\left( E,\,-E^{*} \right)$~~ \\ \hline
		~~chiral~~ & \multicolumn{2}{c}{~~$E=0$~~~or~~~$\left( E,\,-E \right)$~~} \\ \hline
  	\end{tabular}
		\label{tab: spectrum}
\end{table}

We next consider the constraints on complex spectra imposed by anti-unitary symmetry (Table~\ref{tab: spectrum}). A Hamiltonian $H$ has time-reversal and particle-hole symmetries if and only if there exist anti-unitary operators ${\cal T}$ and ${\cal C}$ such that 
\begin{equation}
{\cal T} H {\cal T}^{-1} = H,\quad{\cal C} H {\cal C}^{-1} = - H,
\end{equation}
and ${\cal T}\,z\,{\cal T}^{-1} = z^{*}$, ${\cal C}\,z\,{\cal C}^{-1} = z^{*}$ for all $z \in \mathbb{C}$. For Hermitian Hamiltonians with entirely real spectra, time-reversal symmetry places no constraints on the real spectra and particle-hole symmetry renders the real spectra symmetric about zero energy. By contrast, for non-Hermitian Hamiltonians, of which spectra are not restricted to be real, time-reversal symmetry renders the spectra symmetric about the real axis~\cite{Bender-98}, while particle-hole symmetry makes the spectra symmetric about the imaginary axis~\cite{Malzard-15, Ge-17, KK-tsc18}; 
they are topologically equivalent in the complex energy plane (see Supplementary Note 2 for details). This crucial observation leads to the expectation that non-Hermiticity topologically unifies symmetry classes (Fig.~\ref{fig: AZ symmetry}), as shown below. We note that the role of chiral symmetry is unchanged in non-Hermitian physics, since it is defined to be unitary and does not involve complex conjugation.

\begin{figure}[t]
\centering
\includegraphics[width=50mm]{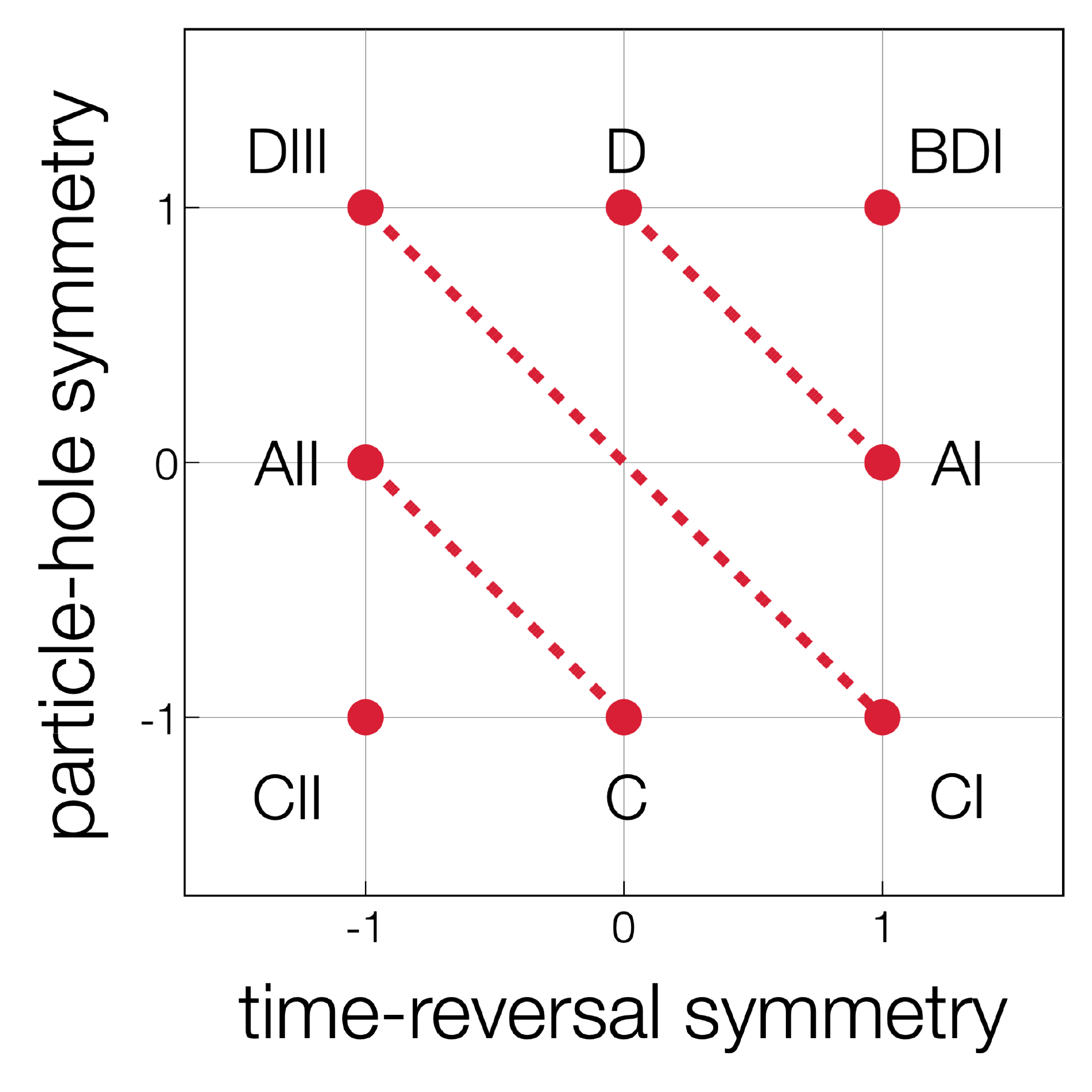} 
\caption{Altland-Zirnbauer symmetry class that involves the anti-unitary symmetry (real class). The classes are specified by the values of ${\cal T}^{2} = \pm 1$ and ${\cal C}^{2} = \pm 1$ and the absence of symmetry is indicated by $0$. The class symbol follows Cartan's notation. While there are eight symmetry classes in Hermitian physics, the classes connected by dotted lines are topologically equivalent and thus the symmetry classes reduce to five in non-Hermitian physics. For instance, a non-Hermitian Hamiltonian having time-reversal symmetry with ${\cal T}^{2} = +1$ alone (class AI) and that having particle-hole symmetry with ${\cal C}^{2} = +1$ alone (class D) are equivalent and hence unified.}
	\label{fig: AZ symmetry}
\end{figure}

\paragraph{Topological unification.\,---}
Motivated by the topological equivalence of time-reversal and particle-hole symmetries in the complex energy plane, we consider a general anti-unitary symmetry ${\cal A}$ defined by Eq.~(\ref{symmetry}). Here ${\cal A}$ reduces to the operator that corresponds to time-reversal (particle-hole) symmetry for $\varphi = 0$ ($\varphi = \pi$). Remarkably, only time-reversal and particle-hole symmetries are allowed when $H$ is Hermitian. To see this, let us take Hermitian conjugation of Eq.~(\ref{symmetry}) and use Hermiticity of $H$ ($H^{\dag} = H$) and the definition of anti-unitary symmetry (${\cal A}^{\dag} = {\cal A}^{-1}$). We then obtain ${\cal A} H {\cal A}^{-1} = e^{-\ii \varphi} H$, which leads to $\varphi = 0,\,\pi$. For non-Hermitian $H$, on the other hand, there are no such constraints.

We study continuous deformations between a system with time-reversal symmetry and a system with particle-hole symmetry in the presence of a complex-energy gap and an anti-unitary symmetry ${\cal A}$. Such deformations cannot be performed for Hermitian Hamiltonians since only the discrete values $\varphi = 0,\,\pi$ are allowed due to Hermiticity; a topological phase with time-reversal symmetry and that with particle-hole symmetry are distinguished in Hermitian physics. Surprisingly, an arbitrary non-Hermitian Hamiltonian $H_{0}$ with time-reversal symmetry can be continuously deformed into a Hamiltonian with particle-hole symmetry because $H_{\varphi} := e^{-\ii \varphi/2} H_{0}$ preserves both complex gap and anti-unitary symmetry ${\cal A}$ for all $\varphi$, and $H_{\pi}$ has particle-hole symmetry. Therefore, a topological phase with time-reversal symmetry and that with particle-hole symmetry are unified into the same topological class in non-Hermitian physics. The topological unification of anti-unitary symmetries presents a general symmetry principle in non-Hermitian physics that holds regardless of the definition of a complex gap~\cite{Gong-18}.

\begin{figure}[t]
\centering
\includegraphics[width=86mm]{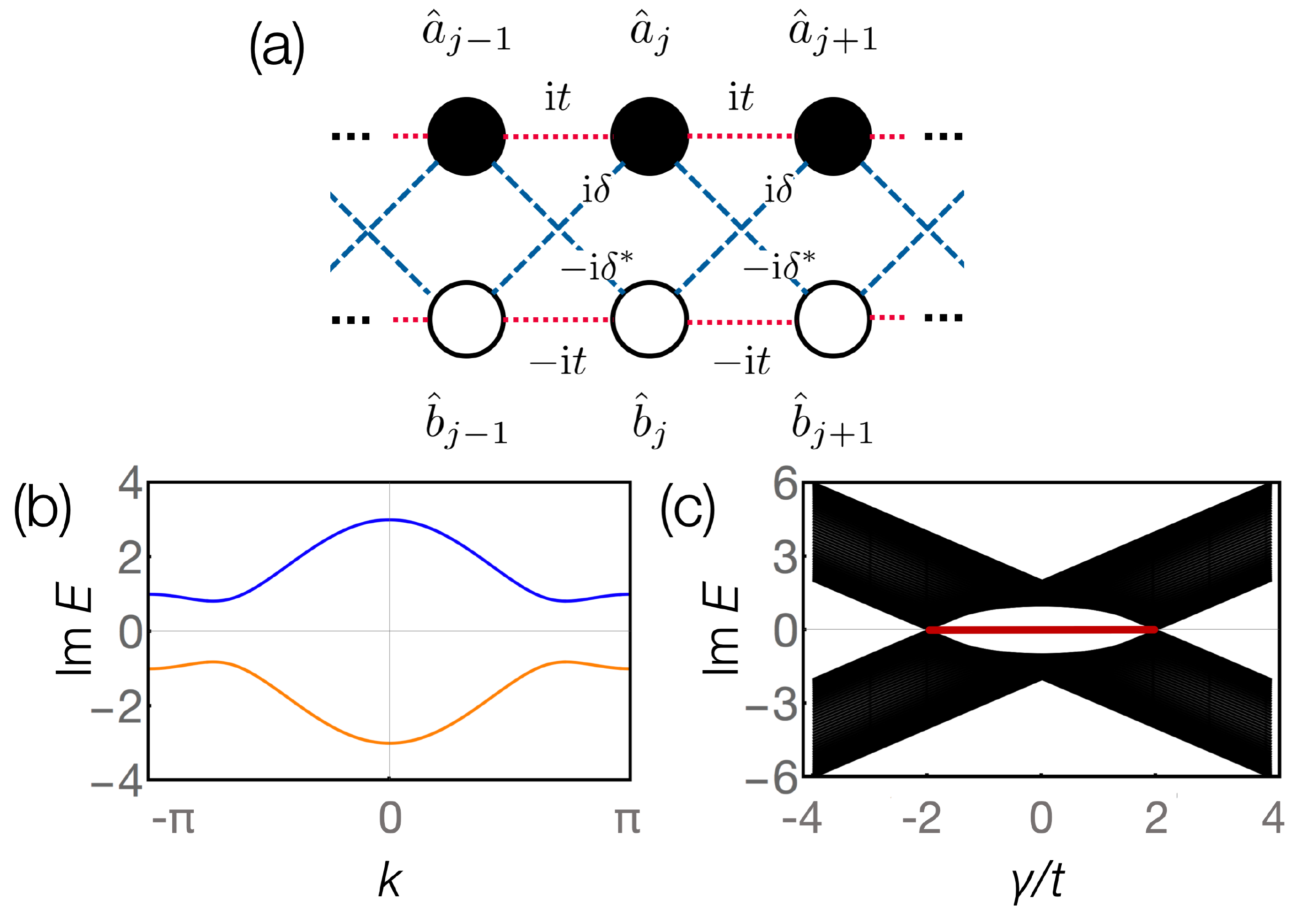} 
\caption{Non-Hermitian one-dimensional topological insulator with time-reversal symmetry (1D class AI) described by Eq.~(\ref{eq: NH-TI}). (a)~Schematic representation of the Hamiltonian. The system consists of the asymmetric hopping and the gain/loss. (b) Energy dispersion of the chain with periodic boundaries ($t=1.0,\,\delta=0.5,\,\gamma=1.0$). The imaginary spectrum is gapped for $\left| \gamma \right| \neq 2t$. (c)~Imaginary spectrum of the system of $L=50$ sites with open boundaries ($t=1.0,\,\delta=0.5$). Edge states with ${\rm Im}\,E=0$ (red line) emerge in the topological phase ($\left| \gamma/t \right| \leq 2$).}
	\label{fig: NH-TI}
\end{figure}

\paragraph{Topological insulator induced by non-Hermiticity.\,---}
As a consequence of the topological unification of time-reversal and particle-hole symmetries, unique non-Hermitian topological phases emerge that are absent in Hermitian systems. In particular, in accordance with the topological phase in the Majorana chain (1D class D)~\cite{Kitaev-01}, non-Hermiticity induces topological phases in one-dimensional insulators that respect time-reversal symmetry with ${\cal T}^{2} = +1$ (1D class AI). Examples include a one-dimensional lattice with two sites per unit cell [Fig.~\ref{fig: NH-TI}\,(a)]:
\begin{eqnarray}
&&\hat{H}_{\rm NHTI}
= \sum_{j} \left\{ \ii t \left( \hat{a}_{j-1}^{\dag} \hat{a}_{j} - \hat{b}_{j-1}^{\dag} \hat{b}_{j} + \hat{a}_{j}^{\dag} \hat{a}_{j-1} - \hat{b}_{j}^{\dag} \hat{b}_{j-1} \right) \right. \nonumber \\
&+& \left[ \ii \delta \left( \hat{b}_{j-1}^{\dag} \hat{a}_{j} - \hat{b}_{j+1}^{\dag} \hat{a}_{j} \right) + \ii \delta^{*} \left( \hat{a}_{j}^{\dag} \hat{b}_{j-1} - \hat{a}_{j}^{\dag} \hat{b}_{j+1} \right) \right] \nonumber \\
&+& \left. \ii \gamma \left( \hat{a}_{j}^{\dag} \hat{a}_{j} - \hat{b}_{j}^{\dag} \hat{b}_{j} \right) \right\},
	\label{eq: NH-TI}
\end{eqnarray}
where $\hat{a}_{j}$ ($\hat{a}_{j}^{\dag}$) and $\hat{b}_{j}$ ($\hat{b}_{j}^{\dag}$) denote the annihilation (creation) operators on each sublattice site $j$; $t > 0$ and $\delta \in \mathbb{C}$ are the asymmetric-hopping amplitudes, and $\gamma \in \mathbb{R}$ is the balanced gain and loss. Such gain and loss have been experimentally implemented in various systems~\cite{Zeuner-15, Weimann-17, Obuse-17, St-Jean-17, Zhou-18, Parto-18, Segev-18, Guo-09, Ruter-10, Regensburger-12, Peng-14, Feng-14, Hodaei-14, Zhen-15, Gao-15, Peng-16, Hodaei-17}, and the asymmetric hopping in optical systems~\cite{Chen-17}. The system respects time-reversal symmetry ($\hat{\cal T}\,\hat{H}_{\rm NHTI}\,\hat{\cal T}^{-1} = \hat{H}$), where the time-reversal operation is defined by $\hat{\cal T}\,\hat{a}_{j}\,\hat{\cal T}^{-1} = \hat{b}_{j}$, $\hat{\cal T}\,\hat{b}_{j}\,\hat{\cal T}^{-1} = \hat{a}_{j}$, and $\hat{\cal T}\,z\,\hat{\cal T}^{-1} = z^{*}$ for all $z \in \mathbb{C}$. The eigenstates form two bands in momentum space, for which the Hamiltonian is determined as $\vec{h} \left( k \right) \cdot \vec{\sigma}$ with $h_{x} = -2\ii\,{\rm Im} \left[ \delta \right] \sin k$, $h_{y} = 2\ii\,{\rm Re} \left[ \delta \right] \sin k$, $h_{z} = \ii \left( \gamma + 2t \cos k \right)$, and Pauli matrices $\vec{\sigma} := ( \sigma_{x},\,\sigma_{y},\,\sigma_{z} )$. The energy dispersion is obtained as $E_{\pm} \left( k \right) = \pm \ii \sqrt{\left( \gamma + 2t \cos k \right)^{2} + 4 \left| \delta \right|^{2} \sin^{2} k}$, and hence the complex bands are separated from each other by the energy gap with magnitude min \{$2 \left| \gamma + 2t \right|$,\,$2 \left| \gamma - 2t \right|$\} [Fig.~\ref{fig: NH-TI}\,(b)].

In parallel with the Majorana chain~\cite{Kitaev-01}, the topological invariant $\nu_{\rm AI}$ is defined by
\begin{equation}
\left( -1 \right)^{\nu_{\rm AI}}
:= {\rm sgn}\,[ \ii h_{z} \left( 0 \right) \cdot \ii h_{z} \left( \pi \right) ]
= -{\rm sgn}\,[ \gamma^{2} - 4t^{2} ].
	\label{eq: topo-inv AI}
\end{equation}
As a hallmark of the non-Hermitian topological phase, a pair of edge states with zero imaginary energy appears when the bulk has non-trivial topology [$\nu_{\rm AI} = 1$; Fig.~\ref{fig: NH-TI}\,(c)]. Whereas the bulk states that belong to the band $E_{+}$ ($E_{-}$) are amplified (attenuated) with time, the mid-gap edge states are topologically protected from such amplification and attenuation. In the case of $t=\delta$, for instance, the topologically protected edge states are obtained as 
\begin{equation} \begin{split}
&\hat{\Psi}_{\rm edge}^{\rm (left)} \propto \ii \sum_{j=1}^{L} \left( - \frac{\gamma}{2t} \right)^{j-1} (\hat{a}_{j} - \hat{b}_{j}), \\
&\hat{\Psi}_{\rm edge}^{\rm (right)} \propto \sum_{j=1}^{L} \left( - \frac{\gamma}{2t} \right)^{j-1} (\hat{a}_{L-j+1} + \hat{b}_{L-j+1}),
\end{split} \end{equation}
which satisfy $\| [\hat{H}_{\rm NHTI},\,\hat{\Psi}_{\rm edge}] \| = O\,(e^{-L/\xi})$ with the localization length $\xi := - ( \log \left| \gamma/2t \right| )^{-1}$. These edge states are immune to disorder that respects time-reversal symmetry (see Supplementary Note 7 for details), which is a signature of the topological phase. We emphasize that topological phases are absent in 1D class AI in the presence of Hermiticity~\cite{Kane-review, Zhang-review, Schnyder-Ryu-review};
non-Hermiticity induces the unique non-equilibrium topological phase as a result of the topological unification of time-reversal and particle-hole symmetries. Whereas the system is an insulator and does not support non-Abelian Majorana fermions, the sublattice degrees of freedom $\hat{a}_{j}$ and $\hat{b}_{j}$ play the roles of particles and holes in the Majorana chain; the Majorana edge states, which are equal superposition states of particles and holes, correspond to the equal superposition states of the two sublattices $\hat{a}_{j}$ and $\hat{b}_{j}$ in the non-Hermitian topological insulator.

\begin{figure}[t]
\centering
\includegraphics[width=78mm]{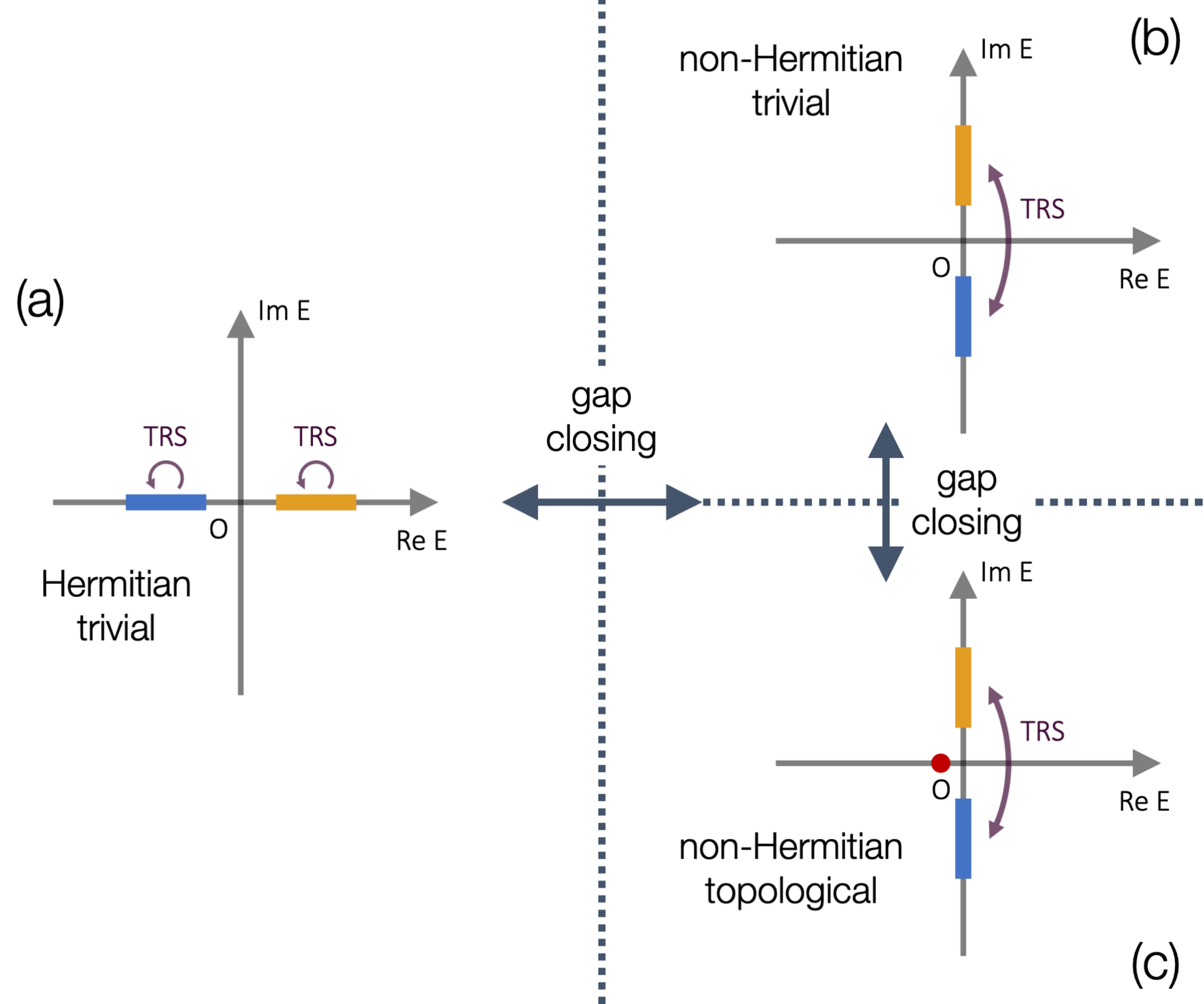} 
\caption{Non-Hermitian one-dimensional insulators with time-reversal symmetry (1D class AI). Blue and yellow bands represent complex bands and red dots represent topologically protected edge states. (a)~Hermitian gapped band structure. All the bands individually respect time-reversal symmetry, and topological phases are absent. (b, c)~Non-Hermitian gapped band structure in 1D class AI for (b)~the trivial phase and (c)~the topological phase. Two bands are paired via time-reversal symmetry and complex-gap closing associated with a topological phase transition should occur between (a) and (b, c).}
	\label{fig: AI}
\end{figure}

\paragraph{Emergent non-Hermitian topological phases.\,---}
The topological phases induced by non-Hermiticity are not specific to the above model but general for all the non-Hermitian systems with anti-unitary symmetry. To see this, we examine the complex-band structure of a generic two-band system $( E_{+} \left( k \right),\,E_{-} \left( k \right) )$ in 1D class AI. In the presence of Hermiticity, the real bands individually respect time-reversal symmetry: $E_{\pm} \left( k \right) = E_{\pm}^{*} \left( -k \right)$ [Fig.~\ref{fig: AI}\,(a)], where topological phases are absent~\cite{Kane-review, Schnyder-Ryu-review}. In the presence of strong non-Hermiticity, on the other hand, time-reversal symmetry is spontaneously broken and the complex bands are paired via time-reversal symmetry: $E_{+} \left( k \right) = E_{-}^{*} \left( -k \right)$. Importantly, this system has the same band structure as the Hermitian topological superconductor protected by particle-hole symmetry (1D class D) as a direct consequence of the topological unification of time-reversal and particle-hole symmetries; it exhibits both trivial [Fig.~\ref{fig: AI}\,(b)] and topological  [Fig.~\ref{fig: AI}\,(c)] phases according to the $\mathbb{Z}_{2}$ topological invariant defined by Eq.~(\ref{eq: topo-inv AI}). The latter band structure becomes gapless in the presence of Hermiticity due to $E_{+} \left( k_{0} \right) = E_{-} \left( k_{0} \right)$ for a time-reversal-invariant momentum $k_{0} \in \{ 0, \pi \}$. 

Remarkably, the emergent non-Hermitian topological phases cannot be continuously deformed into any Hermitian phase that belongs to the same symmetry class. In fact, there should exist a non-Hermitian Hamiltonian that satisfies $E_{+} \left( k \right) = E_{-} \left( -k \right)$ between the two types of band structures and the complex gap closes at $k = k_{0}$. Thus complex-gap closing associated with a topological phase transition should occur between these phases. We also emphasize that the above discussions are applicable to all the non-Hermitian topological phases in any spatial dimension protected by anti-unitary symmetry. Here the corresponding topological invariants are solely determined by the relationship between symmetry and the complex-band structure as in the Hermitian case~\cite{Teo-10}.

\begin{figure}[t]
\centering
\includegraphics[width=80mm]{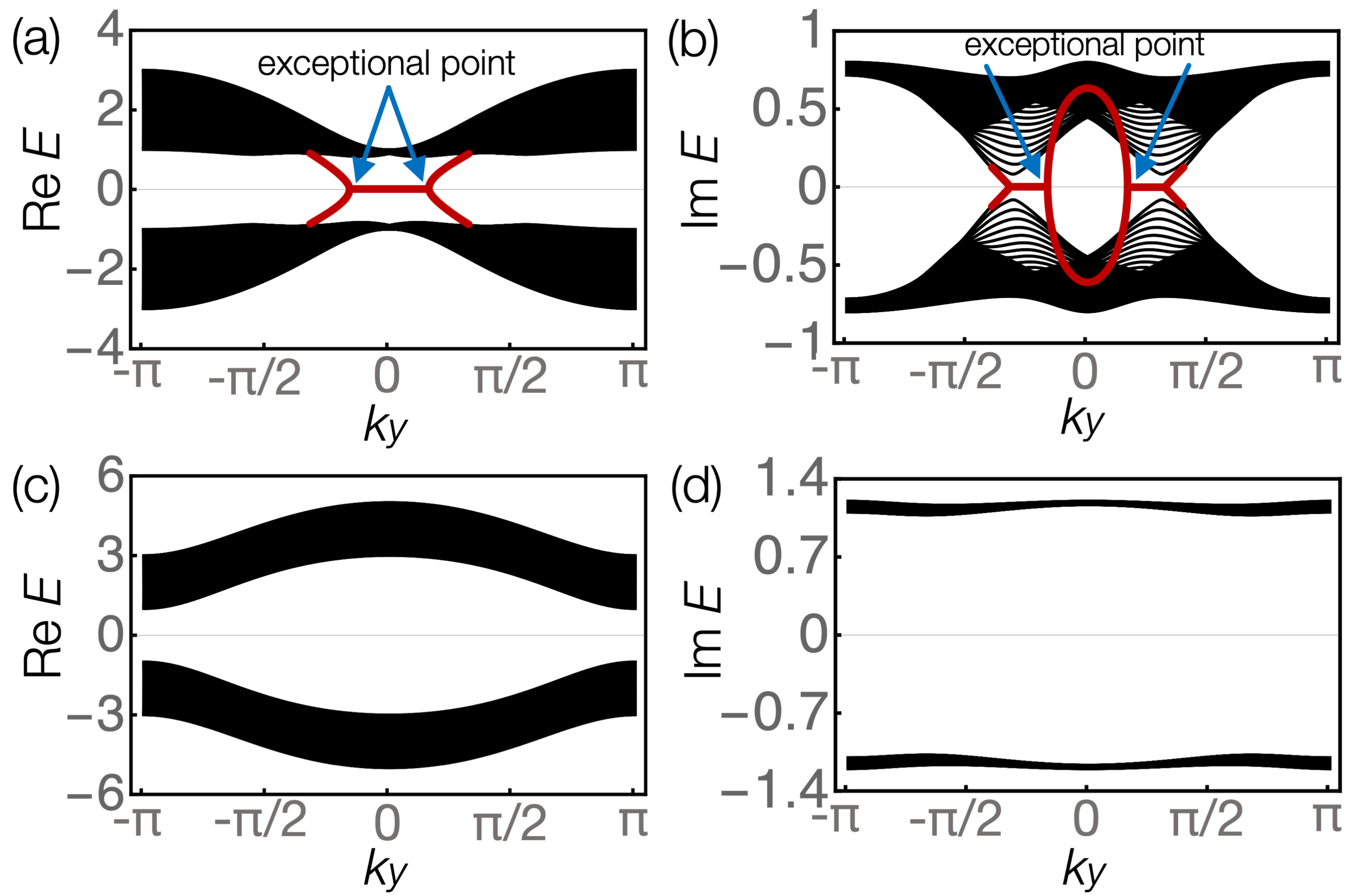} 
\caption{Non-Hermitian quantum spin Hall insulator described by Eq.~(\ref{eq: QSHI}) with $d_{1} \left( {\bm k} \right) = m + t \cos k_{x} + t \cos k_{y}$, $d_{2} \left( {\bm k} \right) = t \sin k_{y}$, $d_{3} \left( {\bm k} \right) = \lambda \left( \sin k_{x} + \sin k_{y} \right)$, $d_{5} \left( {\bm k} \right) = t \sin k_{x}$, and $d_{25} \left( {\bm k} \right) = \gamma$ ($t,\,m,\,\lambda,\,\gamma \in \mathbb{R}$). We here consider a non-Hermitian two-dimensional insulator on a square lattice with open boundaries in the $x$ direction and periodic boundaries in the $y$ direction, along which the wavenumber $k_{y}$ is well-defined. The $\mathbb{Z}_{2}$ topological invariant is given by $\left( -1 \right)^{\nu_{\rm AII}} = {\rm sgn}\,[ m^{2} - 4t^{2} ]$. (a)~Real and (b) imaginary parts of the complex spectrum in the topological phase ($t=1.0$, $m=-1.0$, $\lambda=0.5$, $\gamma=0.8$; $\nu_{\rm AII} = 1$). Helical edge states (red curves) appear between the gapped complex bands and form a pair of exceptional points. (c)~Real and (d) imaginary parts of the complex spectrum in the trivial phase ($t=1.0$, $m=3.0$, $\lambda=0.8$, $\gamma=1.2$; $\nu_{\rm AII} = 0$). No gapless states appear between the gapped complex bands.}
	\label{fig: QSH}
\end{figure}

\paragraph{Quantum spin Hall insulator.\,---}
Topological phases survive non-Hermiticity also in two dimensions. In fact, the $\mathbb{Z}_{2}$ topological invariant $\nu_{\rm AII}$ can be defined in non-Hermitian two-dimensional insulators that respect both time-reversal and parity (inversion) symmetries just like the Hermitian ones~\cite{Fu-07}:
\begin{equation}
\left( -1 \right)^{\nu_{\rm AII}}
:= \prod_{\bm k_{0}} \prod_{n:~{\rm occupied}} \pi_{n} \left( {\bm k_{0}} \right),
	\label{eq: Z2}
\end{equation}
where ${\bm k_{0}} \in \left\{ (0,0),\,(0,\pi),\,(\pi,0),\,(\pi,\pi) \right\}$ denotes the time-reversal-invariant and inversion-symmetric momenta in the Brillouin zone, and $\pi_{n} \left( {\bm k_{0}} \right) \in \left\{ \pm 1 \right\}$ is the parity eigenvalue of the $n$-th Kramers pair at ${\bm k} = {\bm k_{0}}$. In particular, for 4-band insulators such as the Kane-Mele model~\cite{Kane-Mele-05} and the Bernevig-Hughes-Zhang model~\cite{BHZ-06}, the $4 \times 4$ Hamiltonian in momentum space that satisfies ${\cal T}\,{\cal H} \left( {\bm k} \right) {\cal T}^{-1} = {\cal H} \left( -{\bm k} \right)$ and ${\cal P}\,{\cal H} \left( {\bm k} \right) {\cal P}^{-1} = {\cal H} \left( -{\bm k} \right)$ is expressed as
\begin{equation}
{\cal H}_{\rm QSH} \left( {\bm k} \right)
= d_{0} \left( {\bm k} \right) I 
+ \vec{d} \left( {\bm k} \right) \cdot \vec{\Gamma} 
+\ii \sum_{1 \leq i < j \leq 5} d_{ij} \left( {\bm k} \right) \Gamma_{ij},
	\label{eq: QSHI}
\end{equation}
where the coefficients $d_{i}$'s and $d_{ij}$'s are real, $\Gamma_{i}$'s are ${\cal PT}$-symmetric five Dirac matrices, and $\Gamma_{ij}$'s are their commutators $\Gamma_{ij} := [\Gamma_{i},\,\Gamma_{j}]/2\ii$. We notice that Hermiticity leads to $d_{ij} = 0$. Here only $\Gamma_{1}$ and $\Gamma_{ij}$ ($2\leq i<j \leq 5$) are invariant under spatial inversion when $\Gamma_{1}$ is chosen as ${\cal P}$~\cite{Fu-07}. Moreover, when the parity and time-reversal operators are given as ${\cal P} = \sigma_{z}$ and ${\cal T} = \ii s_{y}\,{\cal K}$, the Dirac matrices can be expressed as $\Gamma_{1} = \sigma_{z}\,(= {\cal P})$, $\Gamma_{2} = \sigma_{y}$, $\Gamma_{3} = \sigma_{x} s_{x}$, $\Gamma_{4} = \sigma_{x}  s_{y}$, and $\Gamma_{5} = \sigma_{x} s_{z}$~\cite{Fu-07}. Here $\sigma_{i}$'s ($s_{i}$'s) denote the Pauli matrices that describe the degrees of a sublattice (spin). Since the Hamiltonian at ${\bm k} = {\bm k_{0}}$ is invariant under inversion (${\cal P}\,{\cal H} \left( {\bm k_{0}} \right) {\cal P}^{-1} = {\cal H} \left( {\bm k_{0}} \right)$), it reduces to ${\cal H}_{\rm QSH} \left( {\bm k_{0}} \right) = d_{0} \left( {\bm k_{0}} \right) I + d_{1} \left( {\bm k_{0}} \right) {\cal P} + \ii \sum_{1 \leq i < j \leq 5} d_{ij} \left( {\bm k_{0}} \right) \Gamma_{ij}$; the parity of a Kramers pair at ${\bm k} = {\bm k}_{0}$ corresponds to the sign of $d_{1} \left( {\bm k}_{0} \right)$, and the $\mathbb{Z}_{2}$ topological invariant defined by Eq.~(\ref{eq: Z2}) is obtained as $\left( -1 \right)^{\nu_{\rm AII}} = \prod_{\bm k_{0}} {\rm sgn} \left[ d_{1} \left( {\bm k_{0}} \right) \right]$ as long as complex bands are gapped and $d_{1} \left( {\bm k_{0}} \right)$ is non-zero.

This bulk $\mathbb{Z}_{2}$ topological invariant corresponds to the emergence of helical edge states (Fig.~\ref{fig: QSH}). In stark contrast to Hermitian systems~\cite{Kane-Mele-05, Moore-07, Fu-07, BHZ-06}, the helical edge states form not a Dirac point but a pair of exceptional points~\cite{Xu-17, Shen-18, Zhou-18, Zhen-15, Berry-04, Heiss-12} and have non-zero imaginary energies at the time-reversal-invariant momenta.
Nevertheless, they are immune to disorder due to the generalized Kramers theorem (see Supplementary Notes 5 and 7 for details), which states that all the real parts of energies should be degenerate in the presence of time-reversal symmetry with ${\cal T}^{2} = -1$; the degeneracies of the real parts of energies forbid continuous annihilation of a pair of helical edge states. Notably, the helical edge states are lasing (see Supplementary Note 8 for details) like chiral edge states in a non-Hermitian Chern insulator~\cite{Segev-18}.

The topological unification of anti-unitary symmetry indicates that non-Hermitian systems that respect particle-hole symmetry with ${\cal C}^{2} = -1$ (2D class C) also exhibit the $\mathbb{Z}_{2}$ topological phase, in contrast to the $2\mathbb{Z}$ topological phase in Hermitian physics~\cite{Kane-review, Zhang-review, Schnyder-Ryu-review}.
Here spin-up and spin-down particles in insulators correspond to particles and holes in superconductors. This emergent $\mathbb{Z}_{2}$ topological phase is due to the presence of Kramers pairs of particles and holes with imaginary energies, which are forbidden in Hermitian systems where energies are confined to the real axis; non-Hermiticity brings about topological phases unique to non-equilibrium open systems.

\paragraph{Discussion.\,---}
Non-Hermiticity manifests itself in many disciplines of physics as gain and loss or asymmetric hopping~\cite{Konotop-review, Christodoulides-review}. We have shown that such non-Hermiticity unifies the two fundamental anti-unitary symmetries and consequently topological classification, leading to the prediction of unique non-equilibrium topological phases that are absent at equilibrium. The unveiled topological unification of time-reversal and particle-hole symmetries provides a general symmetry principle in non-Hermitian physics that also justifies a different type of topological classification~\cite{Gong-18}. The modified topological classification implies that the symmetry unification can bring about physics unique to non-Hermitian systems. It merits further study to explore such unusual properties and functionalities that result from our symmetry principle.

This work has explored topological phases characterized by wave functions in non-Hermitian gapped systems, which is a non-trivial generalization of the Hermitian topological phases. By contrast, non-Hermitian gapless systems possess an intrinsic topological structure, which accompanies exceptional points~\cite{Berry-04, Heiss-12} and has no counterparts in Hermitian systems. This topology can be characterized by a complex-energy dispersion~\cite{Lee-16, Xu-17, Shen-18} and distinct from the topology defined by wave functions. Complete theory of non-Hermitian topological systems should be formulated on the basis of these two types of topology in a unified manner, which awaits further theoretical development.

\paragraph{Acknowledgements.\,---}
K.K. thanks Aashish A. Clerk, Li Ge, Hosho Katsura, Masatoshi Sato, Ken Shiozaki, Zhong Wang, and Haruki Watanabe for helpful discussions. This work was supported by KAKENHI Grant No.~JP18H01145, and a Grant-in-Aid for Scientific Research on Innovative Areas ``Topological Materials Science" (KAKENHI Grant No.~JP15H05855) from the Japan Society for the Promotion of Science (JSPS). K.K., S.H., and Y.A. were supported by the JSPS through Program for Leading Graduate Schools (ALPS). S.H. and Y.A. acknowledge support from JSPS (KAKENHI Grant No.~JP16J03619 and No.~JP16J03613). Z.G. acknowledges support from MEXT.





\widetext
\pagebreak

\renewcommand{\theequation}{S\arabic{equation}}
\renewcommand{\thefigure}{S\arabic{figure}}
\renewcommand{\thetable}{S\arabic{table}}
\renewcommand{\bibnumfmt}[1]{[S#1]}
\renewcommand{\citenumfont}[1]{S#1}
\setcounter{equation}{0}
\setcounter{figure}{0}
\setcounter{table}{0}

\section*{Supplementary Note 1: Physical meaning of a complex gap}

The presence of a complex gap has a significant influence on the non-equilibrium wave dynamics. To understand this, we consider a three-level system shown in Supplementary Figure~\ref{fig: imaginary gap}\,(a) as an example. Here the ground state $\ket{g}$ is resonantly coupled to the two excited states $\ket{e_{1}}$ and $\ket{e_{2}}$ that have the same energy levels but the different decay rates (linewidths) $\gamma_{1}$ and $\gamma_{2}$ ($\gamma_{1} < \gamma_{2}$), and the driving strength (Rabi frequency) $\Omega$ between $\ket{g}$ and $\ket{e_{1}}$ is equal to that between $\ket{g}$ and $\ket{e_{2}}$. In the framework of the non-Hermitian band theory, this situation corresponds to the presence of two modes having the same real parts but the different imaginary parts of eigenenergies at a particular wavenumber ${\bm k}$. Then when the wave function of the system is denoted as $\ket{\psi}$, the wave dynamics (in the rotating frame) is governed by 
\begin{equation}
\ii \frac{d}{dt} \ket{\psi} = H \ket{\psi},\quad
H := \frac{1}{2} \left( \begin{array}{@{\,}ccc@{\,}}
      -\ii \gamma_{1} & \Omega & 0 \\
      \Omega & 0 & \Omega \\
      0 & \Omega & - \ii \gamma_{2}
    \end{array} \right),
\end{equation}
where the basis is chosen as $\left( \ket{e_{1}},\,\ket{g},\,\ket{e_{2}} \right)$. Here $\ket{\psi}$ represents an electric-field envelope in classical photonics and a quantum wave function in quantum physics (the Planck constant $\hbar$ is set to unity).

Now let us consider driving the system at the Rabi frequency $\Omega$. When the state is initially prepared to be $\ket{g}$, the intensity or the population for the excited state $\ket{e_{i}}$ ($i=1,2$) at time $t$ is obtained as 
\begin{equation}
p_{i} \left( t \right)
= \left| \braket{e_{i} | e^{-\ii Ht} | g} \right|^{2}.
\end{equation} 
The difference of the decay rates (imaginary parts of the eigenenergies) significantly affects the behavior of the wave dynamics, although the two modes have the same energy levels (real parts of the eigenenergies). In fact, the excitation to the mode with the larger decay rate is suppressed [Supplementary Figure~\ref{fig: imaginary gap}\,(b)]. We remark that such suppression of the wave dynamics due to the presence of dissipation is known as the quantum Zeno effect~\cite{Misra-77, Facchi-02}. Although termed as the quantum Zeno effect due to some historical reasons, this phenomenon occurs in any linear dynamics of probability amplitudes and especially in classical wave dynamics. In fact, an early experimental demonstration was performed for classical light~\cite{Kwiat-95}. These dissipative effects are also relevant in many-body systems~\cite{Syassen-08, Barontini-13}. In particular, even a purely imaginary band gap can play a role similar to that of a real band gap and confine the dynamics of a wave packet within the lower band, provided that the driving strength is weak enough compared with the band gap~\cite{Gong-17}.

\begin{figure}[H]
\centering
\includegraphics[width=120mm]{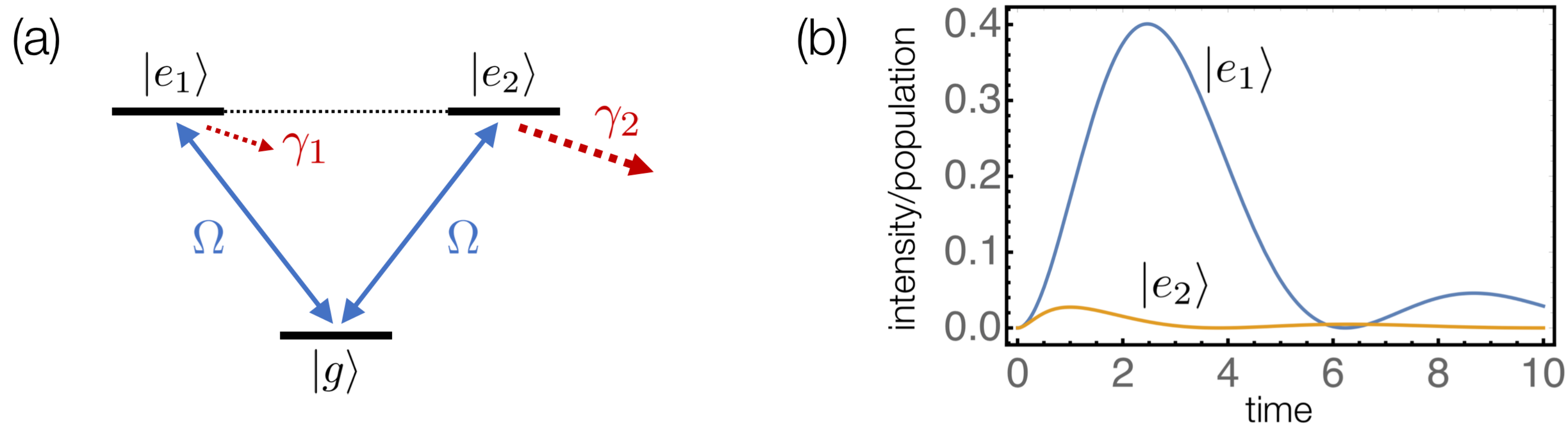} 
\caption{Physical meaning of a complex gap. (a)~Energy-level diagram of a three-level system. The ground state $\ket{g}$ is resonantly coupled to the two excited states $\ket{e_{1}}$ and $\ket{e_{2}}$ that have the same energy levels but the different decay rates (linewidths) $\gamma_{1}$ and $\gamma_{2}$ ($\gamma_{1} < \gamma_{2}$), and the driving strength (Rabi frequency) is equal to $\Omega$ for both excited states. (b)~Evolution of the intensity or the population for the excited states $\ket{e_{1}}$ and $\ket{e_{2}}$ ($\Omega = 1.0,\,\gamma_{1} = 0.5,\,\gamma_{2} = 5.0$). The excitation to the mode $\ket{e_{2}}$ is suppressed due to the larger decay rate ($\gamma_{1} < \gamma_{2}$).}
	\label{fig: imaginary gap}
\end{figure}

\section*{Supplementary Note 2: Symmetry constraints on complex spectra}

We consider a non-Hermitian Hamiltonian $H$ with a generalized antiunitary symmetry $\cal A$ defined by ${\cal A} H {\cal A}^{-1} = e^{\ii \varphi} H$ ($0 \leq \varphi < 2\pi$). When $E \in \mathbb{C}$ is an eigenenergy of $H$ and $\ket{\psi}$ is the corresponding eigenstate, we have
\begin{equation}
H ({\cal A} \ket{\psi})
= e^{-\ii \varphi} {\cal A} H \ket{\psi}
= e^{-\ii \varphi} {\cal A}\,(E \ket{\psi})
= \left( E^{*} e^{-\ii \varphi} \right) ({\cal A} \ket{\psi}).
\end{equation}
Hence ${\cal A} \ket{\psi}$ is an eigenstate of $H$ with eigenenergy $E^{*} e^{-\ii \varphi}$. If $\ket{\psi}$ is also an eigenstate of ${\cal A}$, we have $E = E^{*} e^{-\ii \varphi}$; otherwise eigenenergies come in $(E,\,E^{*} e^{-\ii \varphi})$ pairs. In particular, when ${\cal A}$ describes time-reversal symmetry ($\varphi = 0$), either real eigenenergies or $(E,\,E^{*})$ pairs appear~\cite{Bender-98s}; when ${\cal A}$ describes particle-hole symmetry ($\varphi = \pi$), either pure imaginary eigenenergies or $(E,\,-E^{*})$ pairs appear~\cite{Malzard-15s, Ge-17s, KK-tsc18s}.

\newpage
\section*{Supplementary Note 3: Non-Hermitian Majorana chain}

Certain symmetry-protected topological phases survive even in the presence of non-Hermiticity. Such phases include superconducting wires that respect particle-hole symmetry with ${\cal C}^{2} = +1$ (1D class D). Let us express the Bogoliubov-de Gennes (BdG) Hamiltonian as ${\cal H} \left( k \right) = h_{0} \left( k \right) I + \vec{h} \left( k \right) \cdot \vec{\sigma}$ with the identity matrix $I$ and Pauli matrices $\vec{\sigma} := (\sigma_{x},\,\sigma_{y},\,\sigma_{z})$. Then the presence of particle-hole symmetry ${\cal C} := \sigma_{x} {\cal K}$ implies
\begin{equation}
h_{0, x, y} \left( k \right) = - h_{0, x, y}^{*} \left( -k \right),\quad
h_{z} \left( k \right) = h_{z}^{*} \left( -k \right).
\end{equation}
Therefore, $h_{z}$ becomes real at the particle-hole-symmetric momentum $k_{0} \in \{ 0,\,\pi \}$, and hence the topological invariant $\nu_{\rm D}$ can be defined by 
\begin{equation}
\left( -1 \right)^{\nu_{\rm D}} := {\rm sgn}\,[h_{z} \left( 0 \right) h_{z} \left( \pi \right)]
\end{equation} 
as long as $h_{z} \left( 0 \right)$ and $h_{z} \left( \pi \right)$ are nonzero. The presence of particle-hole symmetry allows one to introduce the topological invariant $\nu_{\rm D}$ as in the Hermitian topological superconductors~\cite{Kitaev-01s}, although $h_{0, x, y} \left( k_{0} \right)$ can be nonzero (pure imaginary) in contrast to Hermitian systems. Moreover, when we introduce Majorana operators $\hat{a}_{k}$ and $\hat{b}_{k}$ in momentum space as 
$\hat{a}_{k}^{\dag} = \hat{a}_{-k}$, $\hat{b}_{k}^{\dag} = \hat{b}_{-k}$, and $\hat{c}_{k} = (\hat{a}_{k}+\ii\hat{b}_{k})/2$,
the Hamiltonian is expressed as 
\begin{equation}
\hat{H}_{\rm D}
= \sum_{k \in {\rm BZ}} \left( \begin{array}{@{\,}cc@{\,}}
      \hat{c}_{k}^{\dag} & \hat{c}_{-k}
    \end{array} \right) \left( \begin{array}{@{\,}cc@{\,}}
      h_{0} + h_{z} & h_{x} -\ii h_{y} \\
      h_{x} + \ii h_{y} & h_{0} -h_{z}
    \end{array} \right) \left( \begin{array}{@{\,}c@{\,}}
      \hat{c}_{k} \\ \hat{c}_{-k}^{\dag}
    \end{array} \right)
= \frac{\ii}{2} \sum_{k \in {\rm BZ}} \left( \begin{array}{@{\,}cc@{\,}}
      \hat{a}_{k}^{\dag} & \hat{b}_{k}^{\dag}
    \end{array} \right) \left( \begin{array}{@{\,}cc@{\,}}
      -\ii h_{0} - \ii h_{x} & \ii h_{y} + h_{z} \\
      \ii h_{y} - h_{z} & - \ii h_{0} + \ii h_{x}
    \end{array} \right) \left( \begin{array}{@{\,}c@{\,}}
      \hat{a}_{k} \\ \hat{b}_{k}
    \end{array} \right).
\end{equation}
Thus, the non-Hermitian part of the momentum-space Hamiltonian at the particle-hole-symmetric momentum $k_{0}$ in the Majorana basis is obtained as $X \left( k_{0} \right) = \ii h_{z} \left( k_{0} \right) \sigma_{y}$, and the topological invariant defined above is also expressed as $\left( -1 \right)^{\nu_{\rm D}} = {\rm sgn}\,[ {\rm Pf}\,X \left( 0 \right) \cdot {\rm Pf}\,X \left( \pi \right) ]$.

The above discussion applies to a non-Hermitian Majorana chain with asymmetric hopping:
\begin{equation} \begin{split}
\hat{H}_{\rm M}
&= \sum_{j} \left[ - t_{L} \hat{c}_{j}^{\dag} \hat{c}_{j+1} - t_{R} \hat{c}_{j+1}^{\dag} \hat{c}_{j}
+ \Delta \hat{c}_{j} \hat{c}_{j+1} +\Delta^{*} \hat{c}_{j+1}^{\dag} \hat{c}_{j}^{\dag} - \mu \left( \hat{c}_{j}^{\dag} \hat{c}_{j} - \frac{1}{2} \right) \right],
	\label{eq: NH-Majorana}
\end{split} \end{equation}
where $\hat{c}_{j}$ ($\hat{c}_{j}^{\dag}$) annihilates (creates) a fermion on site $j$; $t_{L} > 0$ ($t_{R} > 0$) is the hopping amplitude from right to left (from left to right), $\Delta \in \mathbb{C}$ is the $p$-wave pairing gap, and $\mu \in \mathbb{R}$ is the chemical potential. Here non-Hermiticity arises from the asymmetric hopping $t_{L} \neq t_{R}$~\cite{Malzard-15s, Peng-16-pnas, Chen-17s}. The BdG Hamiltonian is given by 
\begin{equation}
h_{0} = \ii \left( t_{L} - t_{R} \right) \sin k,~
h_{x} = 2\,{\rm Im} \left[ \Delta \right] \sin k,~
h_{y} = -2\,{\rm Re} \left[ \Delta \right] \sin k,~
h_{z} = \mu + \left( t_{L} + t_{R} \right) \cos k.
\end{equation} 
Hence the energy dispersion is obtained as 
\begin{equation}
E_{\pm} \left( k \right)
= \ii \left( t_{L} - t_{R} \right) \sin k 
\pm \sqrt{\left( \mu + \left( t_{L} + t_{R} \right) \cos k \right)^{2} + 4 \left| \Delta \right| \sin^{2} k},
\end{equation}
and the complex bands are separated from each other by the gap of magnitude ${\rm min}\,\left\{ 2 \left| \mu + \left( t_{L} + t_{R}\right) \right|,~2 \left| \mu - \left( t_{L} + t_{R}\right) \right| \right\}$ as schematically illustrated in Supplementary Figure~\ref{Majorana}\,(a,b). Moreover, the topological invariant is obtained as $\left( -1 \right)^{\nu_{\rm D}} = {\rm sgn}\,[ \mu^{2} - \left( t_{L} + t_{R} \right)^{2} ]$, and the non-trivial topology of the bulk ($\nu_{\rm D} = 1$; $\left| \mu \right| < t_{L} + t_{R}$) is accompanied by the emergence of the Majorana edge states that are topologically protected with particle-hole symmetry as shown in Supplementary Figure~\ref{Majorana}\,(c,d). Therefore, the topological phase in the Majorana chain survives non-Hermiticity.

\begin{figure}[H]
\centering
\includegraphics[width=172mm]{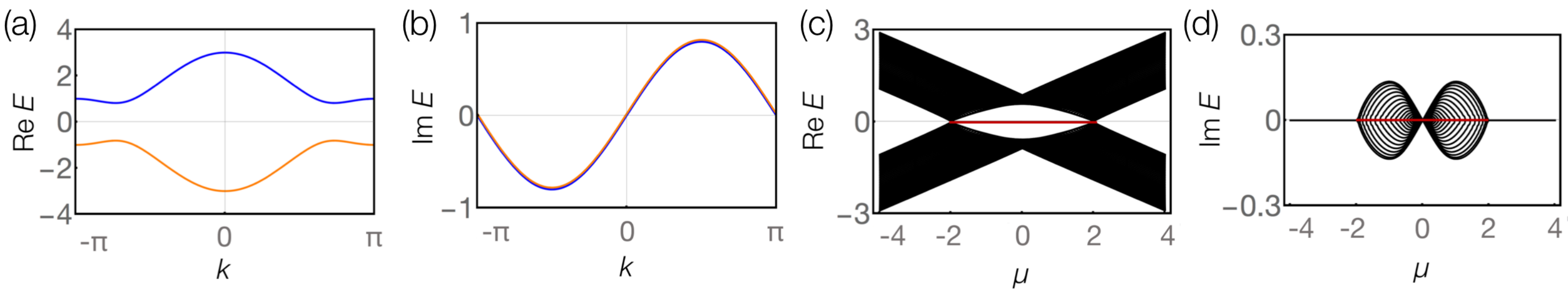} 
\caption{Non-Hermitian Majorana chain with asymmetric hopping (1D class D). (a)~Real and (b)~imaginary parts of the energy dispersion of the chain with periodic boundaries ($t_{L}=1.4,\,t_{R}=0.6,\,\Delta=0.5,\,\mu=1.0$). The complex bands are separated from each other by the gap for $\left| \mu \right| \neq t_{L} + t_{R}$. (c)~Real and (d)~imaginary parts of the complex spectrum of the chain with $50$ sites and open boundaries ($t_{L}=1.4,\,t_{R}=0.6,\,\Delta=0.5$). Majorana edge states with zero energy (red lines) appear in the topological phase ($\left| \mu \right| < t_{L} + t_{R} = 2.0$).}
	\label{Majorana}
\end{figure}

\newpage
\section*{Supplementary Note 4: Symmetry and complex-band structure}

\begin{figure}[b]
\centering
\includegraphics[width=172mm]{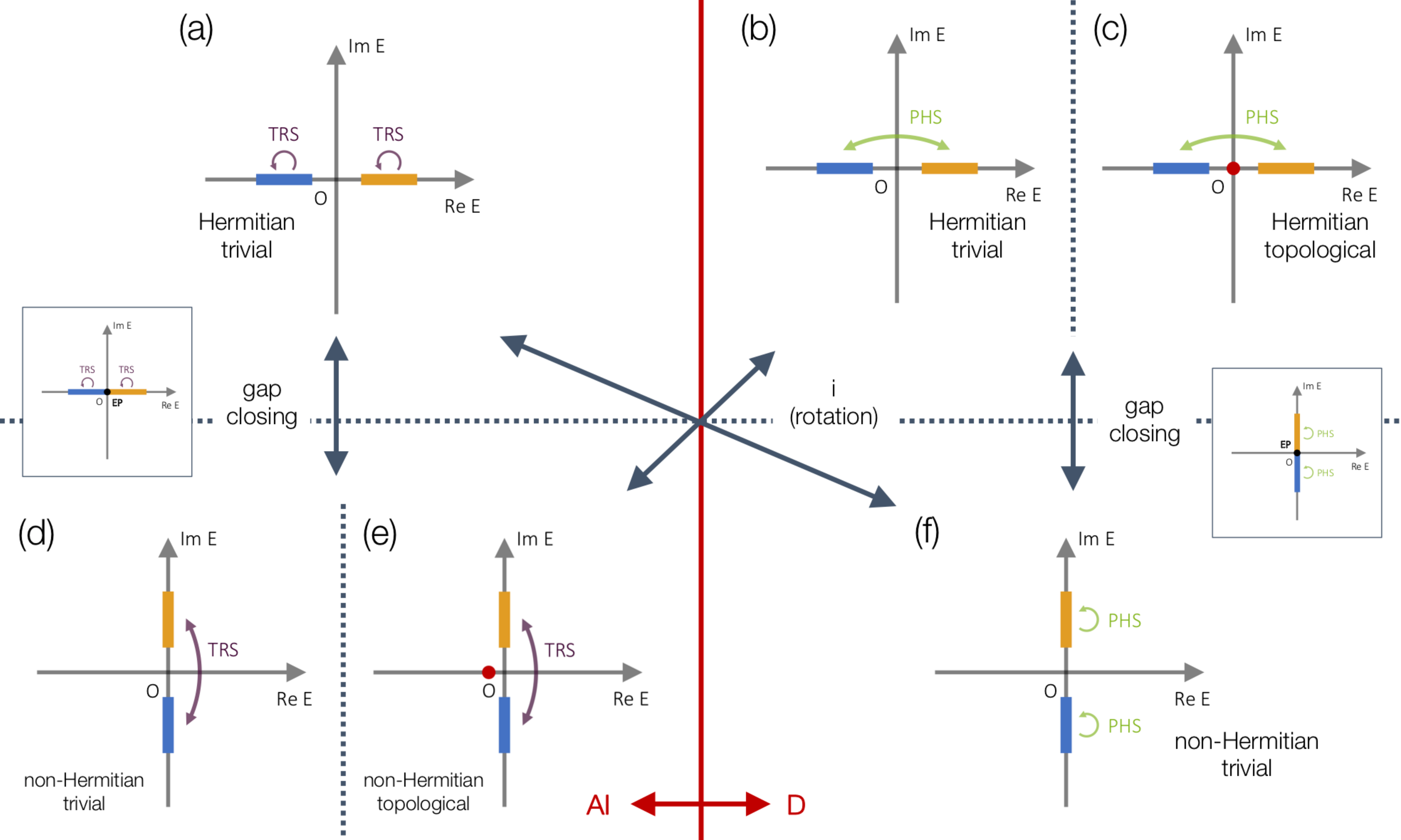} 
\caption{Complex-band structure of one-dimensional systems with time-reversal or particle-hole symmetry (1D class AI or D). Blue and yellow bands represent complex bands and red dots represent the topologically protected edge states. When the system respects time-reversal (particle-hole) symmetry, it belongs to 1D class AI; (a, d, e) [D; (b, e, f)]. (a)~Hermitian gapped band structure in 1D class AI. All the bands individually respect time-reversal symmetry, and topological phases are absent. (b, c)~Hermitian gapped band structure in 1D class D for the (b)~trivial and (c)~topological phases. Two bands are paired via particle-hole symmetry. (d, e)~Non-Hermitian gapped band structure in 1D class AI for the (d)~trivial and (e)~topological phases. Two bands are paired via time-reversal symmetry and gap closing associated with a topological phase transition should occur between (a) and (d, e). (f)~Non-Hermitian gapped band structure in 1D class D. All the bands individually respect particle-hole symmetry, and topological phases are absent.}
	\label{fig: band structure - AI-D}
\end{figure}

\begin{figure}[t]
\centering
\includegraphics[width=110mm]{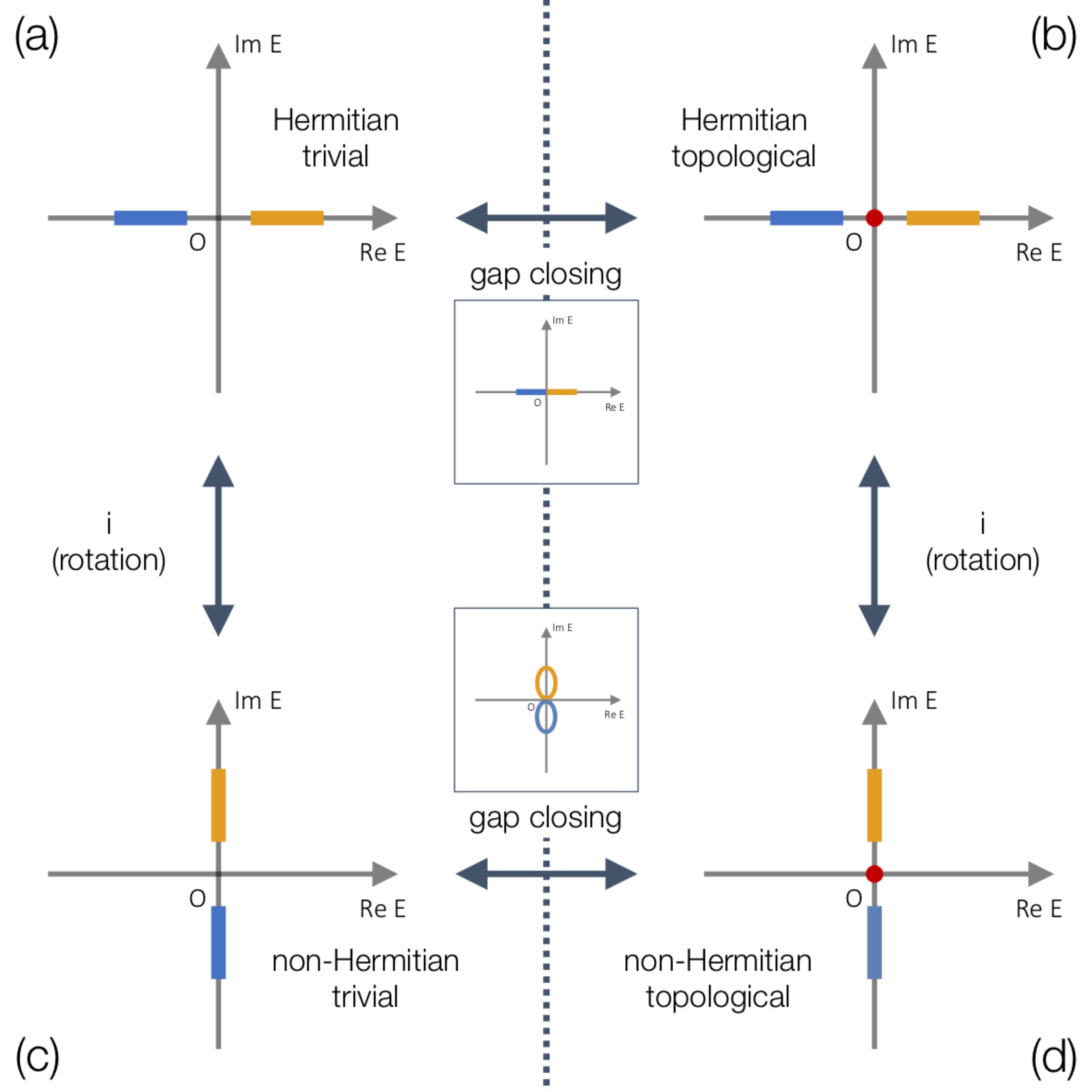} 
\caption{Complex-band structure of one-dimensional systems with chiral symmetry (1D class AIII). Blue and yellow bands represent complex bands that are related to each other via chiral symmetry, and red dots represent the topologically protected states with zero energy. (a)~Hermitian and trivial gapped band structure. (b)~Hermitian and topological band structure. Between (a) and (b), gap closing associated with a topological phase transition should occur. (c)~Non-Hermitian and trivial gapped band structure obtained by multiplying the Hamiltonian in (a) by $\ii$ (rotating the whole complex bands through 90 degrees). (d)~Non-Hermitian and topological gapped band structure. Between (c) and (d), gap closing associated with a topological phase transition should occur. Here (b) and (d) can be continuously deformed to each other.}
	\label{fig: NH-SSH}
\end{figure}

We consider in detail a complex-band structure of a generic one-dimensional system with time-reversal symmetry (1D class AI). We here investigate a two-band system $( E_{+} \left( k \right),~E_{-} \left( k \right) )$ for the sake of simplicity, but the discussion can be straightforwardly generalized to arbitrary $2n$-band systems. In the presence of Hermiticity, the real bands individually respect time-reversal symmetry [Supplementary Figure~\ref{fig: band structure - AI-D}\,(a)]:
\begin{equation}
E_{+} \left( k \right) = E_{+}^{*} \left( -k \right),\quad
E_{-} \left( k \right) = E_{-}^{*} \left( -k \right),
	\label{eq: AI - H}
\end{equation}
where topological phases are absent. In the presence of strong non-Hermiticity, on the other hand, time-reversal symmetry is spontaneously broken and the two complex bands are paired via time-reversal symmetry [Supplementary Figure~\ref{fig: band structure - AI-D}\,(d, e)]: 
\begin{equation}
E_{+} \left( k \right) = E_{-}^{*} \left( -k \right).
	\label{eq: AI - AH}
\end{equation}
Remarkably, this band structure becomes gapless in the presence of Hermiticity since $E_{+} \left( k_{0} \right) = E_{-} \left( k_{0} \right)$ holds for a time-reversal-invariant momentum $k_{0} \in \{ 0,\,\pi \}$, but a complex gap can be open for a non-Hermitian Hamiltonian. This complex-band structure can exhibit both trivial [Supplementary Figure~\ref{fig: band structure - AI-D}\,(d)] and topological [Supplementary Figure~\ref{fig: band structure - AI-D}\,(e)] phases. Between these two types of band structures, there should exist a non-Hermitian Hamiltonian that satisfies both Supplementary Equation~(\ref{eq: AI - H}) and Supplementary Equation~(\ref{eq: AI - AH}) and
\begin{equation}
E_{+} \left( k \right) = E_{-} \left( -k \right).
\end{equation}
Hence the complex gap closes at a time-reversal-invariant momentum $k_{0}$ for this Hamiltonian and gap closing associated with a topological phase transition should be accompanied between these phases [Supplementary Figure~\ref{fig: band structure - AI-D}\,(a, d, e)]. Therefore, the emergent non-Hermitian topological phases cannot be continuously deformed into any Hermitian phase that belongs to the same symmetry class. Importantly, there is a fundamental and non-trivial relationship between symmetry classes AI and D as a direct consequence of the topological unification of time-reversal and particle-hole symmetries. If we begin with a Hermitian Hamiltonian that belongs to 1D class D and possesses the topological phase [Supplementary Figure~\ref{fig: band structure - AI-D}\,(c)] and multiply it by $\ii$, we obtain the non-Hermitian Hamiltonian that belongs to 1D class AI and also possesses the topological phase [Supplementary Figure~\ref{fig: band structure - AI-D}\,(e)] (notice that the eigenstates remain the same under multiplication by $\ii$). The obtained topological phase can be continuously deformed into the emergent non-Hermitian topological phases in 1D class AI.

There is a crucial distinction between this emergent non-Hermitian topological insulator and the non-Hermitian topological insulator protected by chiral symmetry including the Su-Schrieffer-Heeger (SSH) model. In fact, the topological phase in the non-Hermitian SSH model can be continuously deformed into the topological phase in the Hermitian SSH model (Supplementary Figure~\ref{fig: NH-SSH}), whereas those in the topological insulator induced by non-Hermiticity cannot be continuously deformed into any Hermitian phase as discussed above. Here the crucial distinction is that chiral symmetry is unitary (unrelated to complex conjugation) in stark contrast to the anti-unitary time-reversal and particle-hole symmetries (see Table I in the main text).

\section*{Supplementary Note 5: Generalized Kramers theorem}

Anti-unitarity of ${\cal A}$ gives $\braket{\psi | {\cal A} \psi}
= \braket{{\cal A}^{2} \psi | {\cal A} \psi}$. Assuming ${\cal A}^{2} = -1$, we obtain $\braket{\psi | {\cal A} \psi} = - \braket{\psi | {\cal A} \psi}$, which leads to $\braket{\psi | {\cal A} \psi} = 0$. If $\ket{\psi}$ and ${\cal A} \ket{\psi}$ belong to the same eigenenergy ($E = E^{*} e^{-\ii \varphi}$), they are orthogonal and thus degenerate. When ${\cal A}$ describes time-reversal symmetry ($\varphi = 0$), this reduces to the conventional Kramers degeneracy; when ${\cal A}$ describes particle-hole symmetry ($\varphi = \pi$), this implies the degeneracy with pure imaginary eigenenergies.

\section*{Supplementary Note 6: Non-Hermitian quantum spin Hall insulator}

A generic non-Hermitian Hamiltonian that forms four bands in momentum space can be expressed by the identity matrix $I$, five  Dirac matrices $\Gamma_{i}$, and their ten commutators $\Gamma_{ij} := [\Gamma_{i},\,\Gamma_{j}]/2\ii$ as in Eq.~(7) in the main text. If we choose the Dirac matrices to be ${\cal PT}$-symmetric [(${\cal PT})\,\Gamma_{i}\,({\cal PT})^{-1} = \Gamma_{i}$], their commutators become anti-${\cal PT}$-symmetric [(${\cal PT})\,\Gamma_{ij}\,({\cal PT})^{-1} = -\Gamma_{ij}$] due to the anti-unitarity of ${\cal T}$. In the simultaneous presence of time-reversal and inversion symmetries, the Hamiltonian is also ${\cal PT}$-symmetric [(${\cal PT})\,{\cal H}_{\rm QSH} \left( {\bm k} \right)\,({\cal PT})^{-1} = {\cal H}_{\rm QSH} \left( {\bm k} \right)$], which makes the coefficients $d_{0} \left( {\bm k} \right)$, $d_{i} \left( {\bm k} \right)$, and $d_{ij} \left( {\bm k} \right)$ real. Moreover, if we choose $\Gamma_{1}$ as ${\cal P}$, we have~\cite{Fu-07s}
\begin{equation} \begin{split}
{\cal P}\,\Gamma_{i}\,{\cal P}^{-1} = \begin{cases}
+ \Gamma_{1} & {\rm for}~~i=1,\quad \\
- \Gamma_{i} & {\rm for}~~i \neq 1,\quad
\end{cases}&~~~
{\cal T}\,\Gamma_{i}\,{\cal T}^{-1} = \begin{cases}
+ \Gamma_{1} & {\rm for}~~i=1, \\
- \Gamma_{i} & {\rm for}~~i \neq 1;
\end{cases} \\
{\cal P}\,\Gamma_{ij}\,{\cal P}^{-1} = \begin{cases}
- \Gamma_{1j} & {\rm for}~~i=1,\quad \\
+ \Gamma_{ij} & {\rm for}~~i\neq 1,\quad
\end{cases}&~~~
{\cal T}\,\Gamma_{ij}\,{\cal T}^{-1} = \begin{cases}
+ \Gamma_{1j} & {\rm for}~~i=1, \\
- \Gamma_{ij} & {\rm for}~~i \neq 1.
\end{cases}
\end{split} \end{equation}
Therefore, the presence of both time-reversal and inversion symmetries [${\cal T}\,{\cal H}_{\rm QSH} \left( {\bm k} \right)\,{\cal T}^{-1} = {\cal H}_{\rm QSH} \left( -{\bm k} \right)$ and ${\cal P}\,{\cal H}_{\rm QSH} \left( {\bm k} \right)\,{\cal P}^{-1} = {\cal H}_{\rm QSH} \left( -{\bm k} \right)$] implies
\begin{equation}
d_{i} \left( -{\bm k} \right) = \begin{cases}
+ d_{i} \left( {\bm k} \right) & {\rm for}~~i=0,1,\quad \\
- d_{i} \left( {\bm k} \right) & {\rm for}~~i\geq2;\quad
\end{cases}~~~
d_{ij} \left( -{\bm k} \right) = \begin{cases}
- d_{1j} \left( {\bm k} \right) & {\rm for}~~i=1, \\
+ d_{ij} \left( {\bm k} \right) & {\rm for}~~i \geq 2.
\end{cases}
\end{equation}
Hence $d_{i}$'s for $i\geq2$ and $d_{1j}$'s vanish at the time-reversal-invariant and inversion-symmetric momenta ${\bm k} = {\bm k_{0}}$.

\bigskip
\begin{figure}[H]
\centering
\includegraphics[width=160mm]{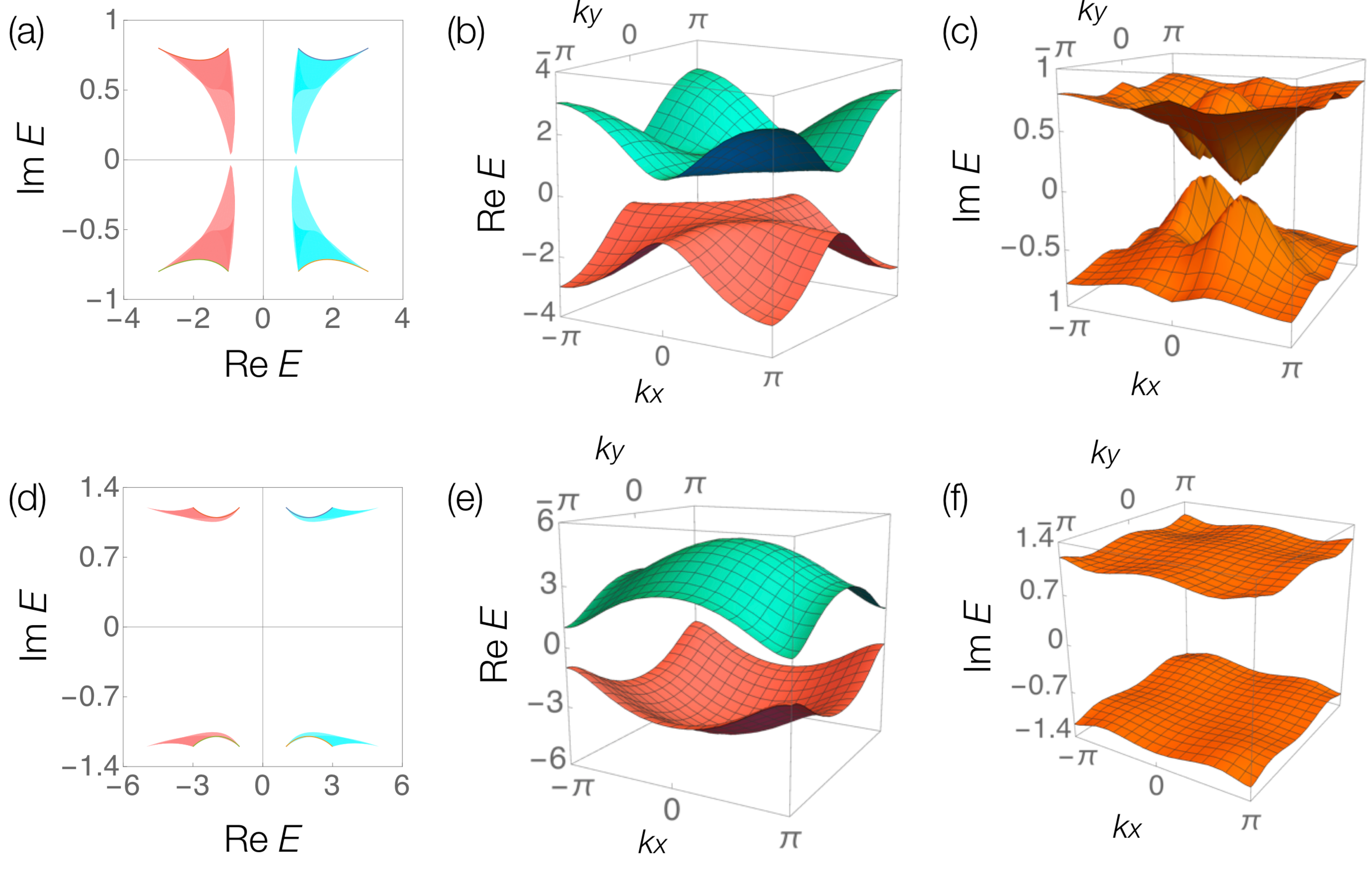} 
\caption{Complex-band structures of the non-Hermitian quantum spin Hall insulator (2D class AII). Cyan (pink) bands correspond to $E \left( k \right),\,E^{*} \left( k \right)$ [$-E \left( k \right),\,-E^{*} \left( k \right)$] in (a, b, d, e). (a)~Gapped bands in the complex energy plane $( {\rm Re}\,E,\,{\rm Im}\,E )$, and (b)~real and (c)~imaginary parts of the energy dispersion as a function of the wavenumber $( k_{x},\,k_{y} )$ in the topological phase ($t=1.0$, $m=-1.0$, $\lambda=0.5$, $\gamma=0.8$; $\nu_{\rm AII} = 1$). (d)~Gapped bands in the complex energy plane $( {\rm Re}\,E,\,{\rm Im}\,E )$, and (e)~real and (f)~imaginary parts of the energy dispersion as a function of the wavenumber $( k_{x},\,k_{y} )$ in the trivial phase ($t=1.0$, $m=3.0$, $\lambda=0.8$, $\gamma=1.2$; $\nu_{\rm AII} = 0$).}
	\label{BHZ-band}
\end{figure}

\newpage
In particular, we consider the non-Hermitian quantum spin Hall insulator described by Eq.~(7) in the main text with
\begin{equation} \begin{split}
d_{1} \left( {\bm k} \right) = m + t \cos k_{x} + t \cos k_{y},~
d_{2} \left( {\bm k} \right) = t \sin k_{y},~
d_{3} \left( {\bm k} \right) = \lambda \left( \sin k_{x} + \sin k_{y} \right),~
d_{5} \left( {\bm k} \right) = t \sin k_{x},~
d_{25} \left( {\bm k} \right) = \gamma,
\end{split} \end{equation}
where $t$, $m$, $\lambda$, and $\gamma$ are real. The eigenstates form four bands in momentum space and their energy dispersions are obtained as $\left( E \left( k \right),\,E^{*} \left( k \right),\,-E \left( k \right),\,-E^{*} \left( k \right) \right)$. Using the anticommutation relations $\{ \Gamma_{i},\,\Gamma_{j} \} = 0$, $\{ \Gamma_{1},\,\Gamma_{25} \} = -2\,\Gamma_{34}$, $\{ \Gamma_{3},\,\Gamma_{25} \} = 2\,\Gamma_{14}$, and $\{ \Gamma_{2},\,\Gamma_{25} \} = \{ \Gamma_{4},\,\Gamma_{25} \} = 0$, we have 
\begin{equation} \begin{split}
{\cal H}_{\rm QSH}^{2} \left( {\bm k} \right)
&= \left[ \left( m+t\cos k_{x} + t \cos k_{y} \right)^{2}  +t^{2} \sin^{2} k_{x} + t^{2} \sin^{2} k_{y} + \lambda^{2} \left( \sin k_{x} + \sin k_{y} \right)^{2} - \gamma^{2} \right] I \\
&~~~~~~~~~~~~~~~~~~~~~~~~~~~~~~~~~~~~~~~~~~~~~~~~~~- 2\ii \gamma \left( m+t\cos k_{x} + t \cos k_{y} \right) \Gamma_{34} 
+ 2\ii \gamma \lambda \left( \sin k_{x} + \sin k_{y} \right) \Gamma_{14}, 
\end{split}\end{equation}
which leads to 
\begin{equation} \begin{split}
E \left( k \right) &= \left\{ \left( m+t\cos k_{x} + t \cos k_{y} \right)^{2}  +t^{2} \sin^{2} k_{x} + t^{2} \sin^{2} k_{y} + \lambda^{2} \left( \sin k_{x} + \sin k_{y} \right)^{2} - \gamma^{2} \right. \\
&~~~~~~~~~~~~~~~~~~~~~~~~~~~~~~~~~~~~~~~~~~~~~~~~~~~~~~~\left. +2\ii \gamma \left[ \left( m+t\cos k_{x} + t\cos k_{y} \right)^{2} + \lambda^{2} \left( \sin k_{x} + \sin k_{y} \right) \right]^{1/2} \right\}^{1/2}.
\end{split} \end{equation}
The complex bands are gapped with the same parameters used in Fig.~5 in the main text (Supplementary Figure~\ref{BHZ-band}). The $\mathbb{Z}_{2}$ topological invariant is determined as
\begin{equation}
\left( -1 \right)^{\nu_{\rm AII}} 
= {\rm sgn}\,\left[ d_{1} \left( 0,0 \right) d_{1} \left( 0,\pi \right) d_{1} \left( \pi,0 \right) d_{1} \left( \pi,\pi \right) \right]
= {\rm sgn}\,\left[ m^{2} - 4 t^{2} \right].
\end{equation}

\section*{Supplementary Note 7: Robustness against disorder}

\begin{figure}[H]
\centering
\includegraphics[width=140mm]{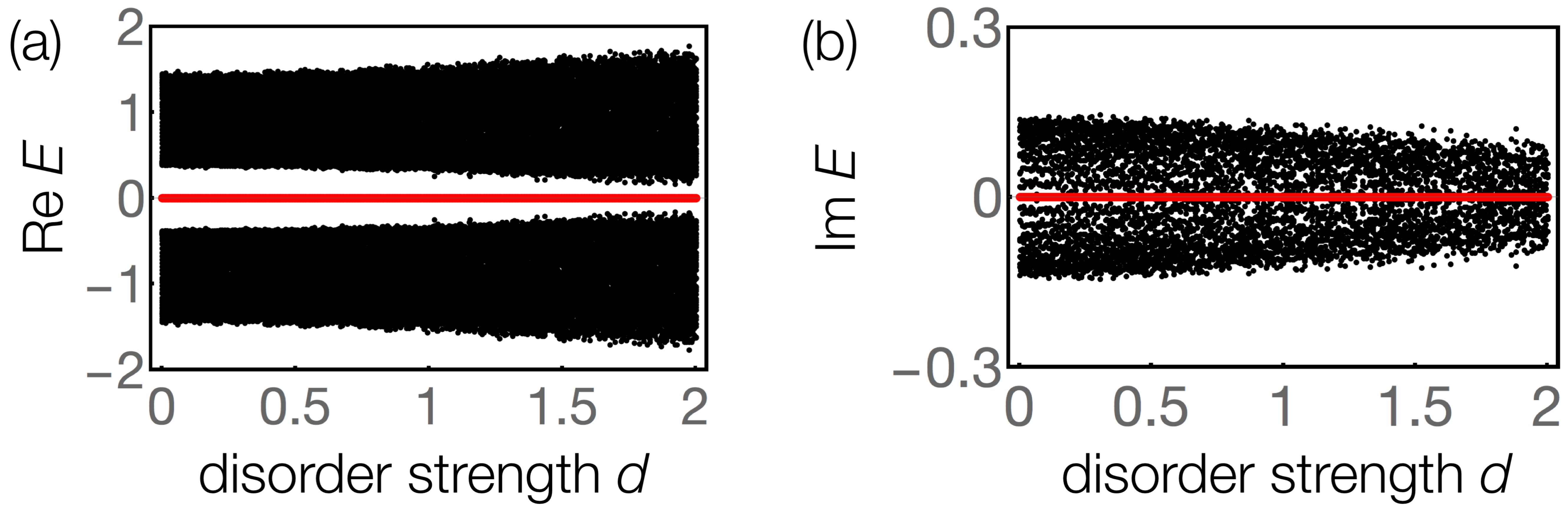} 
\caption{Robustness against disorder in the non-Hermitian Majorana chain. (a)~Real and (b)~imaginary parts of the complex spectrum of the disordered Majorana chain described by Supplementary Equation~(\ref{eq: disordered Majorana}) as a function of the disorder strength $d$. Black and red dots represent the bulk and Majorana edge states, respectively. The chain with $50$ sites is characterized by the parameters $\left( t_{L} \right)_{j} = 1.4+0.3\epsilon_{j}$,~$\left( t_{R} \right)_{j} = 0.6+0.3\epsilon'_{j}$,~$\Delta = 0.5$, and $\mu_{j} = 1.0+d\epsilon''_{j}$, where $\left( t_{L} \right)_{j}$ [$\left( t_{R} \right)_{j}$] is the disordered hopping amplitude from site $j+1$ to $j$ (from site $j$ to $j+1$), $\mu_{j}$ is the disordered chemical potential on site $j$, and $\epsilon_{j}$, $\epsilon'_{j}$ and $\epsilon''_{j}$ are random variables uniformly distributed over $\left[ -0.5,\,0.5 \right]$.}
	\label{NH-Majorana-disorder}
\end{figure}

The topological phase in the non-Hermitian Majorana chain given by Supplementary Equation~(\ref{eq: NH-Majorana}) is immune to disorder that respects particle-hole symmetry. To confirm this robustness, we consider the following disordered non-Hermitian Majorana chain:
\begin{equation}
\hat{H}
= \sum_{j} \left[ - \left( t_{L} \right)_{j} \hat{c}_{j}^{\dag} \hat{c}_{j+1} - \left( t_{R} \right)_{j} \hat{c}_{j+1}^{\dag} \hat{c}_{j} + \Delta \hat{c}_{j} \hat{c}_{j+1} + \Delta^{*} \hat{c}_{j+1}^{\dag} \hat{c}_{j}^{\dag} - \mu_{j} \left( \hat{c}_{j}^{\dag} \hat{c}_{j} - \frac{1}{2} \right) \right],
	\label{eq: disordered Majorana}
\end{equation}
where $\left( t_{L} \right)_{j}$ [$\left( t_{R} \right)_{j}$] is the disordered hopping amplitude from site $j+1$ to $j$ (from site $j$ to $j+1$), and $\mu_{j}$ is the disordered chemical potential on site $j$. The Majorana edge states are protected to have zero energy even in the presence of disorder (Supplementary Figure~\ref{NH-Majorana-disorder}); they are topologically protected with particle-hole symmetry.

\bigskip

\begin{figure}[H]
\centering
\includegraphics[width=140mm]{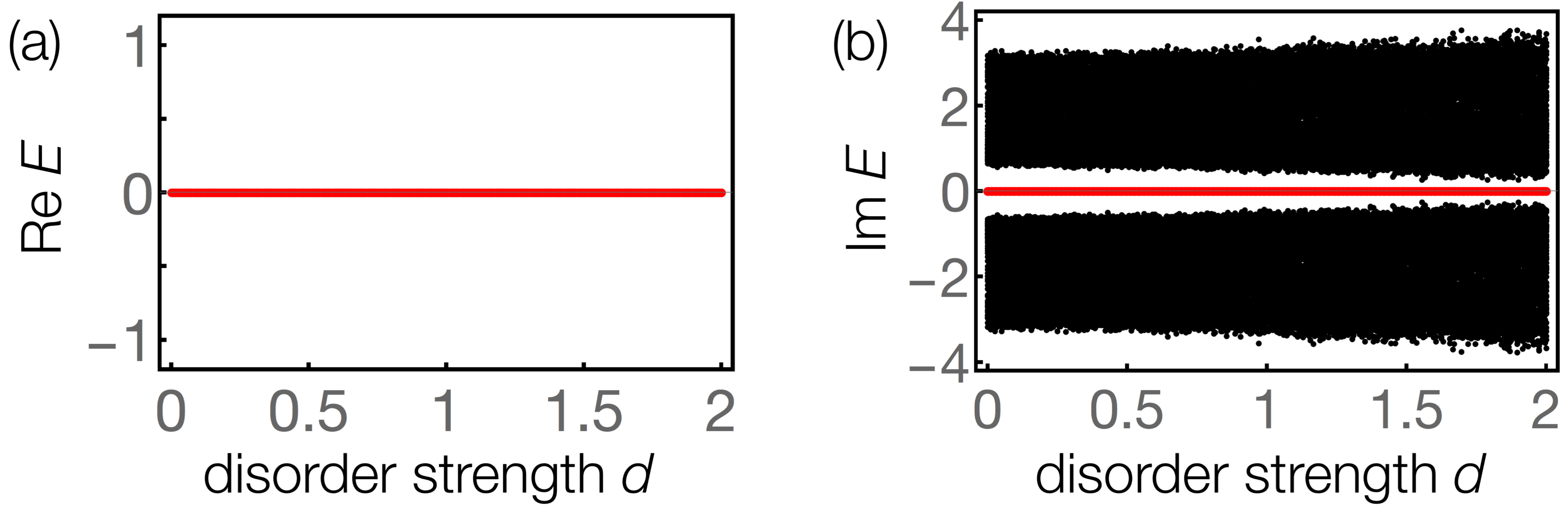} 
\caption{Robustness against disorder in the non-Hermitian topological insulator protected by time-reversal symmetry. (a)~Real and (b)~imaginary parts of the complex spectrum of the disordered non-Hermitian topological insulator described by Supplementary Equation~(\ref{eq: disordered TI}) as a function of the disorder strength $d$. Black and red dots represent the bulk and edge states, respectively. The insulator with $50$ sites is characterized by the parameters $t_{j} = 1.0+0.5\,\epsilon_{j}$,~$\delta_{j} = 0.5 + 0.3\,\epsilon'_{j}$, and $\gamma_{j} = 1.0+d\,\epsilon''_{j}$, where $t_{j}$ and $\delta_{j}$ are the disordered hopping amplitudes between sites $j-1$ and $j$, $\gamma_{j}$ is the disordered gain/loss on site $j$, and $\epsilon_{j}$, $\epsilon'_{j}$, and $\epsilon''_{j}$ are random variables uniformly distributed over $\left[ -0.5,\,0.5 \right]$.}
	\label{NH-TI-disorder}
\end{figure}

The topological phase in the non-Hermitian topological insulator given by Eq.~(3) in the main text is also immune to disorder that respects time-reversal symmetry. To confirm this robustness, we consider the following disordered non-Hermitian chain:
\begin{equation} \begin{split}
\hat{H}
&= \sum_{j} \left[ \ii t_{j} \left( \hat{a}_{j-1}^{\dag} \hat{a}_{j} - \hat{b}_{j-1}^{\dag} \hat{b}_{j} + \hat{a}_{j}^{\dag} \hat{a}_{j-1} - \hat{b}_{j}^{\dag} \hat{b}_{j-1} \right) \right. \\
&~~~~~~~~~~~~~~~~ \left. + \left( \ii \delta_{j} \left( \hat{b}_{j-1}^{\dag} \hat{a}_{j} - \hat{b}_{j+1}^{\dag} \hat{a}_{j} \right) + \ii \delta_{j}^{*} \left( \hat{a}_{j}^{\dag} \hat{b}_{j-1} - \hat{a}_{j}^{\dag} \hat{b}_{j+1} \right) \right) + \ii \gamma_{j} \left( \hat{a}_{j}^{\dag} \hat{a}_{j} - \hat{b}_{j}^{\dag} \hat{b}_{j} \right) \right],
	\label{eq: disordered TI}
\end{split} \end{equation}
where $t_{j}$ and $\delta_{j}$ are the disordered hopping amplitudes between sites $j-1$ and $j$, and $\gamma_{j}$ is the disordered gain/loss on site $j$. The edge states are protected to have zero energy even in the presence of disorder (Supplementary Figure~\ref{NH-TI-disorder}); they are topologically protected with time-reversal symmetry.

\bigskip

\begin{figure}[H]
\centering
\includegraphics[width=172mm]{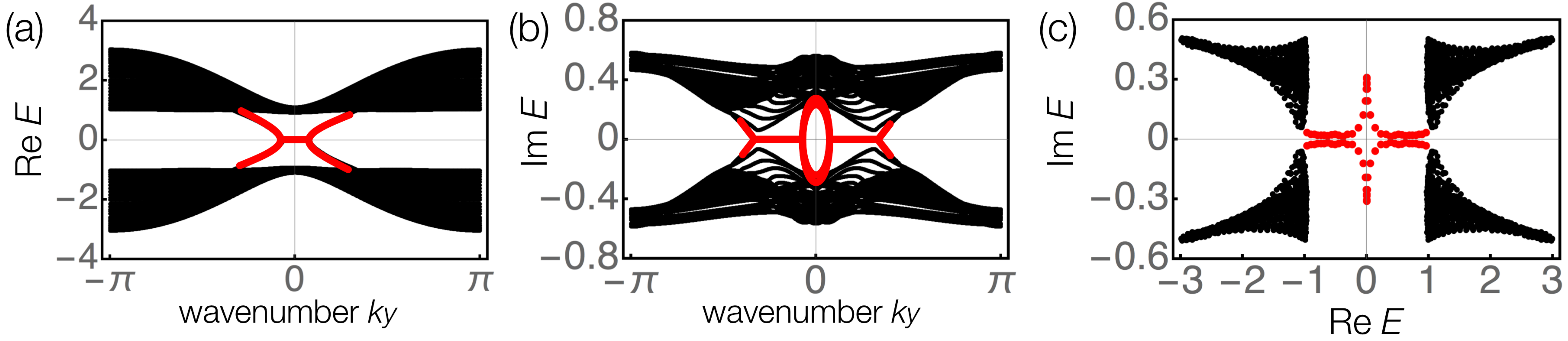} 
\caption{Robustness against disorder in the non-Hermitian quantum spin Hall insulator. (a)~Real and (b)~imaginary parts of the complex spectrum of the disordered non-Hermitian quantum spin Hall insulator described by Supplementary Equation~(\ref{eq: disordered QSH}) as a function of the wavenumber $k_{y}$. Black and red dots represent the bulk and edge states, respectively. The open boundary condition is imposed in the $x$ direction, whereas the periodic boundary condition is imposed in the $y$ direction. The insulator with $30$ sites is characterized by the parameters $t=1.0$, $m_{x} = -1.0+0.3\,\epsilon_{x}$, $\lambda_{x} = 0.5+0.2\,\epsilon'_{x}$and $\gamma_{x} = 0.5 +0.3\,\epsilon''_{x}$, where $\epsilon_{x}$, $\epsilon'_{x}$, and $\epsilon''_{x}$ are random variables uniformly distributed over $\left[ -0.5,\,0.5 \right]$. (c)~Complex spectrum of the disordered non-Hermitian quantum spin Hall insulator described by Supplementary Equation~(\ref{eq: disordered QSH xy-1}) and Supplementary Equation~(\ref{eq: disordered QSH xy-2}). The open boundary condition is imposed in both $x$ and $y$ directions. The insulator with $30 \times 30$ sites is characterized by the parameters $t=1.0$, $m_{x,y} = -1.0+0.3\,\epsilon_{x,y}$, $\lambda_{x,y} = 0.5+0.2\,\epsilon'_{x,y}$ and $\gamma_{x,y} = 0.5 +0.3\,\epsilon''_{x,y}$.}
	\label{NH-QSH-disorder}
\end{figure}

Finally, the topological phase in the non-Hermitian quantum spin Hall insulator given by Eq.~(7) and shown in Fig.~5 in the main text is immune to disorder that respects time-reversal symmetry. To confirm this robustness, we consider the following disordered non-Hermitian insulator with open boundaries in the $x$ direction and periodic boundaries in the $y$ direction:
\begin{equation} \begin{split}
\hat{H} = \sum_{x} \sum_{k_{y} \in {\rm BZ}}
&\left\{ \frac{1}{2} \left[ \hat{\vec{c}}_{x,\,k_{y}}^{~\dag} \left( t \left( \sigma_{z} \otimes I - \ii \sigma_{x} \otimes s_{z}\right) - \ii \lambda_{x} \sigma_{x} \otimes s_{x} \right) \hat{\vec{c}}_{x+1,\,k_{y}} + {\rm H.c.} \right] \right. \\
&~~~~~~~~\left. + \hat{\vec{c}}_{x,\,k_{y}}^{~\dag} \left[ \left( m_{x} + t \cos k_{y} \right) \sigma_{z} \otimes I + \left( t \sin k_{y} \right) \sigma_{y} \otimes I + \left( \lambda_{x} \sin k_{y} \right) \sigma_{x} \otimes s_{x} - \ii \gamma_{x} \sigma_{z} \otimes s_{z} \right] \hat{\vec{c}}_{x,\,k_{y}} \right\},
	\label{eq: disordered QSH}
\end{split} \end{equation}
where $\hat{\vec{c}}_{x,\,k_{y}}$ ($\hat{\vec{c}}_{x,\,k_{y}}^{~\dag}$) annihilates (creates) a fermion on site $x$ and with momentum $k_{y}$ that has four internal degrees of freedom; $m_{x}$, $\lambda_{x}$, and $\gamma_{x}$ denote the disordered parameters. The helical edge states are topologically protected with time-reversal symmetry even in the presence of disorder [Supplementary Figure~\ref{NH-QSH-disorder}\,(a,b)]. Moreover, we confirm their robustness also in the following disordered non-Hermitian insulator with open boundaries in both $x$ and $y$ directions:
\begin{equation} \begin{split}
\hat{H} = \sum_{x} \sum_{y}
\left\{ \frac{1}{2} \left[ 
\hat{\vec{c}}_{x,\,y}^{~\dag}\,T^{\,(x)}_{x,\,y}\,\hat{\vec{c}}_{x+1,\,y}
+ \hat{\vec{c}}_{x,\,y}^{~\dag}\,T^{\,(y)}_{x,\,y}\,\hat{\vec{c}}_{x,\,y+1}
+ {\rm H.c.} \right] 
+ \hat{\vec{c}}_{x,\,y}^{~\dag}\,M_{x,\,y}\,\hat{\vec{c}}_{x,\,y} \right\},
	\label{eq: disordered QSH xy-1}
\end{split} \end{equation}
with 
\begin{equation}
\left( \begin{array}{@{\,}c@{\,}}
      T^{\,(x)}_{x,\,y} \\ T^{\,(y)}_{x,\,y}
    \end{array} \right)
:= \left( \begin{array}{@{\,}c@{\,}}
      t \left( \sigma_{z} \otimes I - \ii \sigma_{x} \otimes s_{z}\right) - \ii \lambda_{x,y}\,\sigma_{x} \otimes s_{x}\\ 
      t \left( \sigma_{z} \otimes I - \ii \sigma_{y} \otimes I\right) - \ii \lambda_{x,y}\,\sigma_{x} \otimes s_{x}
    \end{array} \right),~~
M_{x,\,y} := m_{x,y}\,\sigma_{z} \otimes I - \ii \gamma_{x,y}\,\sigma_{z} \otimes s_{z}.
	\label{eq: disordered QSH xy-2}
\end{equation}
Here $\hat{\vec{c}}_{x,\,y}$ ($\hat{\vec{c}}_{x,\,y}^{~\dag}$) annihilates (creates) a fermion on site ($x$, $y$), and $m_{x,y}$, $\lambda_{x,y}$, and $\gamma_{x,y}$ denote the disordered parameters. The helical edge states are robust also in this case [Supplementary Figure~\ref{NH-QSH-disorder}\,(c)]; the topological phase in the non-Hermitian quantum spin Hall insulator is immune to disorder that respects time-reversal symmetry.

\section*{Supplementary Note 8: Dynamics of topological edge states in non-Hermitian systems: \\ experimental signature of non-Hermitian topological phases}

\begin{figure}[b]
\centering
\includegraphics[width=110mm]{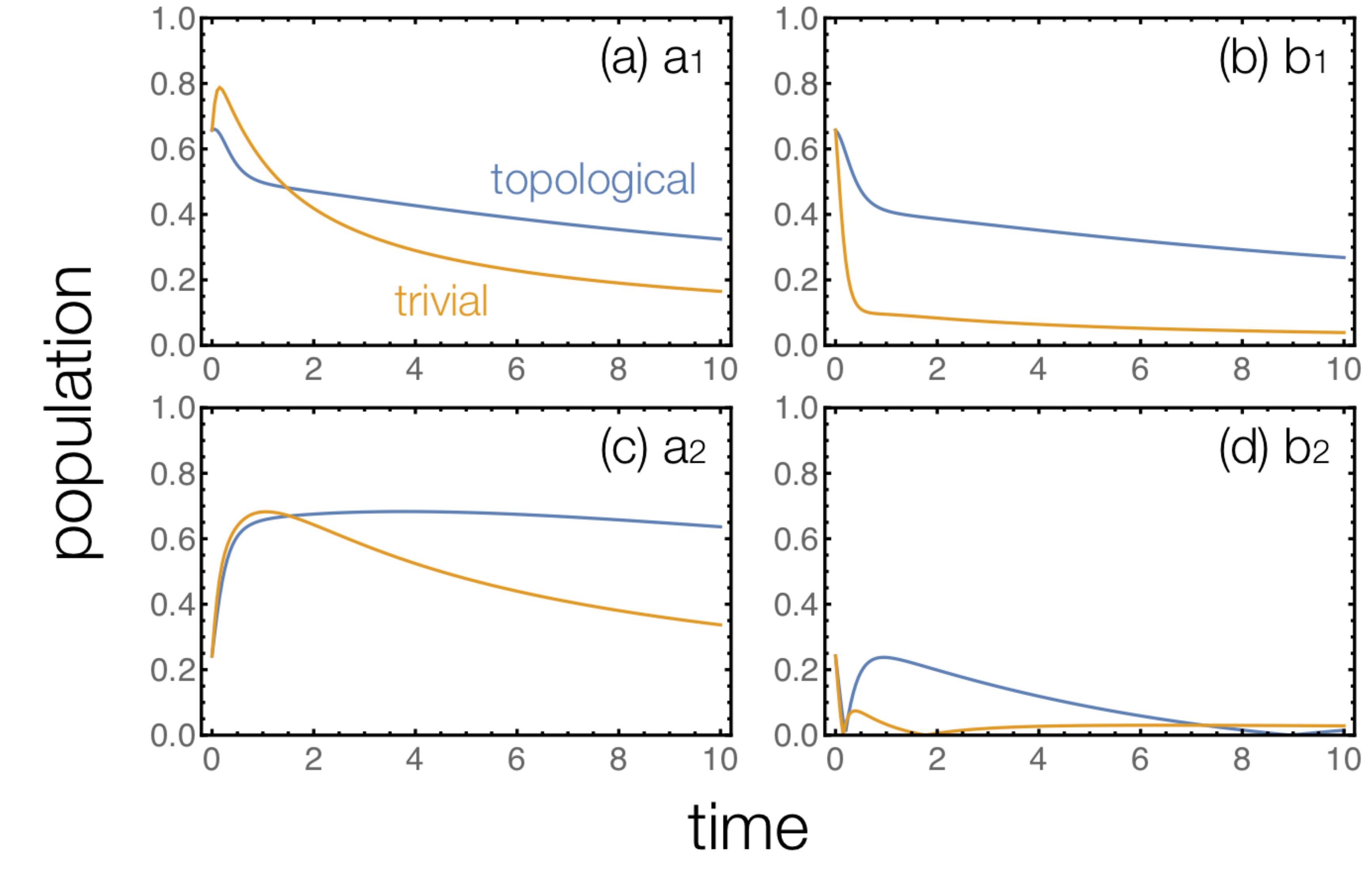} 
\caption{Population dynamics in the non-Hermitian topological insulator protected by time-reversal symmetry (1D class AI). An initial state is prepared to be a localized wave function $\ket{\psi_{0}} \propto \sum_{x=1}^{L} e^{-x} \ket{x}$, and the evolutions of the population (normalized intensity) $\left| \braket{x|\psi_{t}} \right|^{2}/\braket{\psi_{t}|\psi_{t}}$ are shown for (a)~$\ket{x} = \ket{a_{1}}$, (b)~~$\ket{x} = \ket{b_{1}}$, (c)~~$\ket{x} = \ket{a_{2}}$ and (d)~~$\ket{x} = \ket{b_{2}}$ [see Fig. 3\,(a) in the main text for details]. The insulator with $L = 50$ sites is characterized by the parameters $t = \delta = 1.0$ and $\gamma =0.2$ for the topological phase (blue curves) and $\gamma = 3.0$ for the trivial phase (orange curves). In the topological phase, there emerges time-reversal-symmetry-protected topological edge states whose imaginary parts vanish. As a consequence, the population of the wave function in the topological phase is greater than that in the trivial phase.}
	\label{fig: edge - 1D - AI}
\end{figure}

\begin{figure}[t]
\centering
\includegraphics[width=172mm]{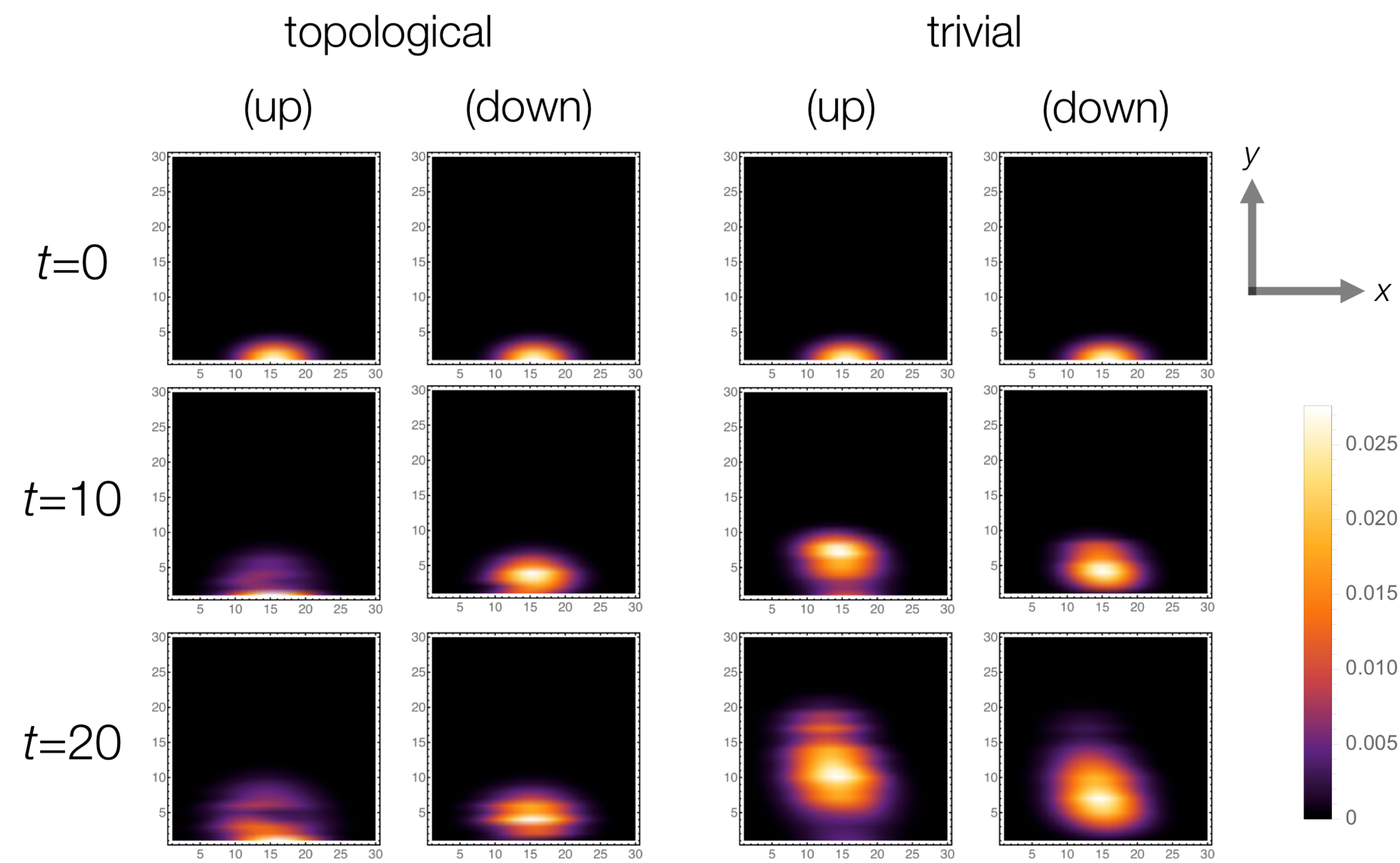} 
\caption{Population dynamics in the non-Hermitian quantum spin Hall insulator (2D class AII). The insulator with $30 \times 30$ sites is characterized by the parameters $t = 1.0,\,m=-1.0,\,\lambda=0.5,\,\gamma=0.8$ for the topological phase (left) and $t = 1.0,\,m=3.0,\,\lambda=0.8,\,\gamma=1.2$ for the trivial phase (right). An initial state is a localized wave function $\ket{\psi_{0}} \propto \sum_{x,y} e^{- \left( x- 15.5 \right)^{2}/36 - \left( y - 1\right)^{2}/9} \ket{x, y}$, and the evolutions of the population (normalized intensity) $\left| \braket{x,y |\psi_{t}} \right|^{2}/\braket{\psi_{t}|\psi_{t}}$ are shown at $t=0,\,10,\,20$. In the topological phase, the wave packet remains localized due to the presence of the helical edge states. In contrast to the Hermitian quantum spin Hall insulator, the spin-up component is more localized than the spin-down component due to the presence of the imaginary gap between the helical edge states. Moreover, the wave packet does not move along the edges since the real parts of the energy dispersions are flat for the helical edge states. In the trivial phase, on the other hand, the wave packet quickly diffuses into the bulk due to the absence of the edge states.}
	\label{fig: edge - 2D - AII}
\end{figure}

A hallmark of the topological phases is the emergence of robust edge states that reflect the non-trivial topology of the bulk. Remarkably, non-Hermiticity can make these edge states amplified (lasing)~\cite{Obuse-17s, St-Jean-17s, Parto-18s, Segev-18s} and anomalous~\cite{Lee-16s}. The non-Hermitian topological insulators considered in this work also exhibit unique topologically protected edge states that have no Hermitian counterparts, which serves as an experimental signature of the non-Hermitian topological phases. In the following, we consider the dynamics of non-Hermitian topological systems. When an initial state is prepared to be
\begin{equation}
\ket{\psi_{0}}
= \sum_{n} c_{n} \ket{\varphi_{n}},\quad
c_{n} = \braket{\chi_{n} | \psi_{0}},
\end{equation}
where $\ket{\varphi_{n}}$ ($\ket{\chi_{n}}$) is a right (left) eigenstate of the non-Hermitian Hamiltonian~\cite{Brody-16}, the wave function evolves as
\begin{equation}
\ket{\psi_{t}}
= e^{-\ii H t} \ket{\psi_{0}}
= \sum_{n} c_{n} e^{-\ii E_{n} t} \ket{\varphi_{n}}.
\end{equation}
Although the system eventually reaches a stationary state $\ket{\varphi_{n}}$ with the largest imaginary part of $E_{n}$, unique topological features appear in the non-Hermitian transient dynamics, which have been observed in recent experiments~\cite{Obuse-17s, St-Jean-17s, Parto-18s, Segev-18s}.

For the non-Hermitian topological insulator given by Eq.~(3) in the main text, the imaginary parts of the eigenenergies of the topologically protected edge states vanish due to the presence of time-reversal symmetry. Consequently, these edge states are stable, despite instability of the bulk states. The presence of such stable edge states can be detected by focusing on the evolution of the particle population near the edges. In fact, when an initial state is prepared as a localized state at one edge, its population near the edge in the topological phase is greater than that in the trivial phase (Supplementary Figure~\ref{fig: edge - 1D - AI}). 

For the non-Hermitian quantum spin Hall insulator given by Eq.~(7) and shown in Fig.~5 in the main text, the helical edge states emerge corresponding to the non-trivial topology of the bulk. As a consequence, when an initial state is prepared as a localized state at one edge, the wave packet remains localized near the edge in the topological phase (Supplementary Figure~\ref{fig: edge - 2D - AII}). In the trivial phase, on the other hand, the wave packet quickly diffuses into the bulk due to the absence of the edge states. There are several significant distinctions between the Hermitian and non-Hermitian quantum spin Hall insulators. In contrast to the Hermitian one, the spin-up component is more localized than the spin-down component due to the presence of the imaginary gap between the helical edge states [see Fig.~4\,(b) in the main text]. Moreover, the wave packet does not move along the edges since the real parts of the energy dispersions are flat for the helical edge states [see Fig.~4\,(a) in the main text].

\section*{Supplementary Note 9: Bulk-edge correspondence in non-Hermitian systems}

The bulk-edge correspondence in Hermitian topological systems is generally understood with a continuum model described by a Dirac Hamiltonian~\cite{Bernevig-textbook, Asboth-textbook}. Although the topological classification is drastically altered by non-Hermiticity as demonstrated in the main text, the bulk-edge correspondence can be generalized to a non-Hermitian continuum model in the same manner as the Hermitian one. We here consider a one-dimensional Dirac Hamiltonian that respects chiral symmetry ${\cal S} := \sigma_{z}$ (hence belongs to 1D class AIII):
\begin{equation}
H \left( k \right)
= \left( k+\ii g \right) \sigma_{x} 
+ \left( m+\ii \delta \right) \sigma_{y},
\end{equation}
where $k$ is a wavenumber (momentum), and $g,\,m,\,\delta \in \mathbb{R}$ are real parameters that characterize the topological or trivial phase. The energy dispersion of this continuum model is obtained as
\begin{equation}
E \left( k \right)
= \pm \sqrt{\left( k+\ii g \right)^{2} + \left( m+\ii \delta \right)^{2}}
= \pm \sqrt{k^{2} + m^{2} - g^{2} - \delta^{2} + 2 \ii \left( kg + m\delta \right)},
\end{equation}
and hence the complex bands are gapped for $\left| m \right| > \left| g \right|$. As with the Hermitian case, the topological invariant (winding number) can be defined for the gapped phases as 
\begin{equation}
W = \int_{-\infty}^{\infty} \frac{dk}{4\pi \ii}\,{\rm tr}\left[ {\cal S} H^{-1} \frac{dH}{dk} \right]
= -\int_{-\infty}^{\infty} \frac{dk}{2\pi} \frac{m+\ii \delta}{\left( k+\ii g \right)^{2} + \left( m+\ii \delta\right)^{2}}
= - \frac{{\rm sgn} \left[ m \right]}{2}.
\end{equation}
Here the half-integer topological invariant $W$ is common to the continuum models and should be complemented by the structure of wave functions away from the Dirac point~\cite{Haldane-88s, Schnyder-08s}.

To investigate the emergence of topologically protected boundary states, we consider a domain wall defined by
\begin{equation}
g \left( x \right) = g_{+} \theta \left( x \right) + g_{-} \theta \left( -x \right),\quad
m \left( x \right) = m_{+} \theta \left( x \right) + m_{-} \theta \left( -x \right),\quad
\delta \left( x \right) = \delta_{+} \theta \left( x \right) + \delta_{-} \theta \left( -x \right),
\end{equation} 
with a step function $\theta \left( x \right)$. Then we solve the Schr\"odinger equation
\begin{equation}
\left[ \left( - \ii \frac{d}{dx} + \ii g \left( x \right) \right) \sigma_{x} + \left( m \left( x \right) + \ii \delta \left( x \right) \right) \sigma_{y} \right] \varphi \left( x \right) = 0
\end{equation}
to find the topologically protected boundary states with zero energy. For $x>0$, if we take an ansatz $\varphi_{+} \left( x \right) = \left( a~b \right)^{T} e^{-x/\xi_{+}}$, the above Schr\"odinger equation reduces to
\begin{equation}
\left( \xi_{+}^{-1} + g_{+} + m_{+} + \ii \delta_{+} \right) a = \left( \xi_{+}^{-1} + g_{+} - m_{+} - \ii \delta_{+} \right) b = 0.
\end{equation}
We notice that ${\rm Re} \left[ \xi_{+} \right] > 0$ is needed for the localization of the boundary states. Then, for $m_{+} > 0$, we have $g_{+} + m_{+} > 0$ and hence ${\rm Re} \left[ \xi_{+}^{-1} + g_{+} + m_{+} + \ii \delta_{+} \right] > 0$; for $m_{+} < 0$, we have $g_{+} - m_{+} > 0$ and hence ${\rm Re} \left[ \xi_{+}^{-1} + g_{+} - m_{+} - \ii \delta_{+} \right] > 0$. Therefore, we obtain
\begin{equation} \begin{split}
\varphi_{+} \left( x \right)
&= \left( \begin{array}{@{\,}c@{\,}}
      0 \\ 1
    \end{array} \right) \exp \left[ -\left( m_{+} - g_{+} + \ii \delta_{+} \right) x \right]~~{\rm for}~~m_{+} > 0;\\
\varphi_{+} \left( x \right)
&= \left( \begin{array}{@{\,}c@{\,}}
      1 \\ 0
    \end{array} \right) \exp \left[ -\left( -m_{+} - g_{+} - \ii \delta_{+} \right) x \right]~~{\rm for}~~m_{+} < 0.
\end{split} \end{equation}
For $x < 0$, on the other hand, we obtain
\begin{equation} \begin{split}
\varphi_{-} \left( x \right)
&= \left( \begin{array}{@{\,}c@{\,}}
      1 \\ 0
    \end{array} \right) \exp \left[ -\left( m_{-} + g_{-} + \ii \delta_{-} \right) x \right]~~{\rm for}~~m_{-} > 0;\\
\varphi_{-} \left( x \right)
&= \left( \begin{array}{@{\,}c@{\,}}
      0 \\ 1
    \end{array} \right) \exp \left[ -\left( -m_{-} + g_{-} - \ii \delta_{-} \right) x \right]~~{\rm for}~~m_{-} < 0.
\end{split} \end{equation}
Due to the boundary condition $\varphi_{+} \left( 0 \right) = \varphi_{-} \left( 0 \right)$, the bound states emerge if and only if the signs of $m_{+}$ and $m_{-}$ are different, reflecting the different winding numbers in each region; the emergence of the topologically protected bound states corresponds to the non-trivial topology of the bulk. Notably, similar discussions for non-Hermitian continuum models in two dimensions are found in Supplementary References~\cite{Leykam-17s, Shen-18s}.

\newpage

\end{document}